\documentclass[a4paper,11pt]{article}
\pdfoutput=1 
\usepackage{jheppub}
\usepackage[utf8]{inputenc}
\usepackage{mathrsfs}		
\usepackage{braket}			
\usepackage{tensor}
\usepackage{comment}

\def\hre#1#2{\href{http://arxiv.org/abs/#1/#2}{[ArXiv:#1/#2]}}
\def\hri#1#2{\href{http://arxiv.org/abs/#1}{[ArXiv:#1]#2}}

\def\a{\alpha}		\def\b{\beta}		\def\g{\gamma}		\def\d{\delta}
\def\e{\varepsilon}			\def\h{\eta}		\def\q{\theta}
				\def\l{\lambda}		\def\m{\mu}
						\def\p{\pi}			\def\r{\rho}	
\def\s{\sigma}						
\def\c{\chi}		\def\w{\omega}			
\def\ve{\e}
\def\G{\Gamma}		\def\D{\Delta}				
			\def\P{\Pi}

		\newcommand{\calL}{{\mathcal L}}
		\newcommand{\calO}{{\mathcal O}}

			\renewcommand{\>}{\rangle}

\DeclareMathOperator{\Tr}{Tr}

\DeclareMathOperator{\Res}{Res}

\newcommand{\der}{{\partial}}
\newcommand{\idnty}{{\hbox{1$\!\!$1}}}
\newcommand{\ap}{{\alpha^\prime}}	

\newcommand{\ie}{{\it i.e.\ }}

\title{\boldmath On the N-pion extension \\ of the Lovelace-Shapiro model}

\preprint{NORDITA-2020-018,\,\, UWThPh-2020-6}

\author[a,b]{Massimo Bianchi}
\author[a,b,c]{Dario Consoli} 
\author[d,e]{Paolo Di Vecchia}

\affiliation[a]{Dipartimento di Fisica, Universit{\`{a}} di Roma ``Tor Vergata'', \\ Via della Ricerca Scientifica,  00133 Roma, Italy}
\affiliation[b]{I.N.F.N. Sezione di Roma ``Tor Vergata'', Via della Ricerca Scientifica, \\ 00133 Roma, Italy}
\affiliation[c]{Mathematical Physics Group, University of Vienna, Boltzmanngasse 5 1090 Vienna, Austria}
\affiliation[d]{The Niels Bohr Institute, University of Copenhagen, Blegdamsvej 17, \\ DK-2100 Copenhagen, Denmark}
\affiliation[e]{Nordita, KTH Royal Institute of Technology and Stockholm University, \\Roslagstullsbacken 23, SE-10691 Stockholm, Sweden}

\emailAdd{massimo.bianchi@roma2.infn.it}
\emailAdd{dario.consoli@univie.ac.at}
\emailAdd{divecchi@nbi.dk}

\abstract{We reconsider a modification of the $N$-point amplitude of the Neveu-Schwarz (NS) model in which the tachyon becomes a pion by shifting its mass to zero and keeping the super-projective invariance of the integrand of the amplitude. For the scattering of four particles it reduces to the amplitude written by Lovelace and Shapiro that has Adler zeroes. We confirm that also the $N$-pion amplitude has Adler zeroes and show that it reduces to that of the non-linear $\sigma$-model for $\alpha' \rightarrow 0$ keeping $F_\pi$ fixed. The four- and six-point flavour-ordered amplitudes satisfy tree-level unitarity since they can be derived from the correspondent amplitudes of the NS model in ten dimensions by suitably choosing the components of the momenta of the external mesons in the six extra dimensions.
Negative norm states (ghosts) are shown to appear instead  in higher-point amplitudes. We also discuss several amplitudes involving different external mesons.}

\begin{document} 
\maketitle
\flushbottom

\section{Introduction}
\label{sec:intro}

Soon after Veneziano's proposal for the $\pi\pi\pi\omega$ scattering amplitude~\cite{Veneziano:1968yb},
a model for $\pi \pi$ scattering was proposed by Lovelace~\cite{LOVELACE} and Shapiro~\cite{SHAPIRO}~\footnote{See also Ref. \cite{FRAMPTON}.}.
According to this model the three isospin amplitudes for pion-pion scattering are given by: 
\begin{eqnarray}
 A_{I=0} =\frac{3}{2} \left( A(s,t)+ A (s,u) \right)- \frac{1}{2} A (t,u)~;~A_{I=1} = A(s,t) - A(s,u)~;~A_{I=2} =A(t,u)~~
\label{I3}
\end{eqnarray}
where
\begin{eqnarray}
A(s,t) = C_\textup{4-pt} \frac{\Gamma (1- \alpha_\rho (s)) \Gamma (1- \alpha_\rho(t) )}{\Gamma (1- \alpha_\rho (s)- \alpha_\rho (t))} \quad{\rm with} \quad  \alpha_\rho (s) = \alpha_0 + \alpha' s 
\label{I4}
\end{eqnarray}
and $C_\textup{4-pt}$ is a constant fixed by the $\ap \to 0$ limit of the amplitude.
The $\rho$ Regge trajectory $\alpha_\rho (s)$ was determined by requiring the Adler zeroes and imposing $\alpha_\rho (s=m^2_\rho) =1$. Taking for simplicity massless pions one gets:
\begin{equation}
\alpha_\rho (s) = \frac{1}{2} + \alpha' s~~;~~ \alpha' = \frac{1}{2m_\rho^2}
\label{I5}
\end{equation}
In the meanwhile, the Veneziano model was extended to the scattering of four  and then of $N$ scalar particles\footnote{For a complete set of references concerning these early developments see Ref.~\cite{PDV}.} of mass $m$, related to the intercept of the Regge trajectory by the relation $\alpha_0 + \alpha' m^2=0$. Written in terms of Koba-Nielsen variables $z_i$ the $N$-point amplitude takes the form:
\begin{eqnarray}
A^{VM}_N(k_i) = \int \frac{\prod_{i=1}^N dz_i}{dV_{abc}} \prod_{i<j} (z_i - z_j)^{2\alpha' k_i k_j} \prod_{i=1}^N (z_i - z_{i+1})^{-\alpha' m^2-1}
\label{I1}
\end{eqnarray}
where $z_1 >z_2 \dots >z_N$ and $dV_{abc} = \frac{dz_a dz_b dz_c}{(z_a - z_b)(z_a- z_c)(z_b-z_c)}$. The invariance of the integrand under projective transformations allows one to fix three variables and integrate over the remaining $N{-}3$ variables. A convenient choice is $z_1 = \infty, z_2 =1$ and $z_N=0$.  Introducing vertex operators $V(k_i , z_i ) =  {\rm e}^{i k_i X(z_i)}$ for the external scalar particles, the previous amplitude can be written as 
\begin{eqnarray}
A^{VM}_N(k_i) = \int \frac{\prod_{i=1}^N dz_i}{dV_{abc}}  \prod_{i=1}^N (z_i - z_{i+1})^{-\alpha' m^2-1}
\langle 0 | \prod_{i=1}^N V (k_i ; z_i ) |0\rangle  
\label{I2}
\end{eqnarray}
Another generalisation of the Veneziano model was the Ramond-Neveu-Schwarz model that not only allowed to include external fermions~\cite{PR}  but also provided a more realistic model for $\pi$ mesons. The scattering amplitude for $N$ mesons with mass $m$ was written as follows~\cite{NS}:
\begin{eqnarray}
A^{NS}_N = \int \frac{\prod_{i=1}^N dz_i}{dV_{abc}}  \prod_{i=1}^N (z_i - z_{i+1})^{-\alpha' m^2- \frac{1}{2}}
\langle 0 | \prod_{i=1}^N k_i \psi (z_i)  {\rm e}^{i k_i X(z_i)} |0 \rangle
\label{I6}
\end{eqnarray}
where $\psi^{\mu} (z)$ are free world-sheet fermions.
As already noticed  in Ref.~\cite{NS}, in order to have all the gauge conditions necessary to eliminate the negative norm states (ghosts), it is necessary to require that the pion is actually a tachyon with a mass given by $m^2 = -\frac{1}{2\alpha'}$. The NSR model has then critical dimension $D=10$. After GSO projection, that alas eliminates the tachyon/pion, the NSR model becomes the superstring theory.

For the previous value of the tachyon mass the NS model has been reformulated in Ref.~\cite{NST}
and the $N$-point tachyon amplitude has been written as follows:
\begin{eqnarray}
A^{NS}_N = \int \prod_{i=2}^{N-1} dz_i \langle 0, k_1| \prod_{i=2}^{N-1} k_i \psi (z_i) {\rm e}^{i k_i X(z_i)} |0, k_N \rangle
\label{I8}
\end{eqnarray}
where one can fix the M{\"{o}}bius invariance by choosing $z_1 =\infty , z_2 =1$ and $z_N=0$. This amplitude can be rewritten in a more compact way by resorting to a super-conformal formalism in two dimensions whereby one associates an anticommuting variable $\theta$ with the Koba-Nielsen variable $z$ and defines a superconformal field ${\cal X}(z, \theta) = X(z) + \theta \psi (z)$ that is a  function of both $z$ and $\theta$~\cite{FM,BW,KH}. Introducing the superconformal vertex operator:
\begin{equation}
V_{NS} (z, \theta; k) = {\rm e}^{i k {\cal X} (z, \theta)}
\label{I9}
\end{equation}
one can write the amplitude in Eq. (\ref{I8}) as follows~\cite{FM,BW,KH}:
\begin{eqnarray}
A_N^{NS} =  \int \frac{\prod_{1=1}^N   dZ_i}{dV_{abc}} \langle 0| \prod_{i=1}^N {\rm e}^{i k_i {\cal X}(z_i, \theta_i)} |0 \rangle  =  \int \frac{\prod_{1=1}^N d \theta_i  dz_i}{d V_{abc}} \prod_{i<j} (Z_i - Z_j )^{2\alpha' k_i k_j}
\label{I10}
\end{eqnarray}
where $\alpha' k_i^2 = \frac{1}{2}$ and 
\begin{equation}
Z_i - Z_j \equiv  z_i -z_j - \theta_i \theta_j ~~;~~ dZ_i \equiv dz_i d \theta_i~~;~~
dV_{abc} = \frac{dZ_a dZ_b dZ_c}{\left[ (Z_a-Z_b) (Z_a-Z_c)(Z_b -Z_c)\right]^{\frac{1}{2}}} 
\frac{1}{d \Theta}
\label{I10a}
\end{equation}
$\Theta$ is the super-projective invariant variable:
\begin{eqnarray}
&&\Theta= \theta_c \left[\frac{Z_a -Z_b}{(Z_a - Z_c) (Z_b -Z_c) }\right]^{\frac{1}{2}} - \theta_b
\left[\frac{Z_a -Z_c}{(Z_a - Z_b) (Z_b -Z_c) }\right]^{\frac{1}{2}}  \nonumber \\
&& + \theta_a \left[\frac{Z_b -Z_c}{(Z_a - Z_b) (Z_a -Z_c) }\right]^{\frac{1}{2}}  - \frac{\theta_c \theta_b \theta_a}{\left[ (Z_a-Z_b) (Z_a-Z_c)(Z_b -Z_c)\right]^{\frac{1}{2}} }
\label{I10b}
\end{eqnarray}
In deriving (\ref{I10}) one has to use the contraction rule
 \begin{equation}
\langle{\cal X}^\mu(Z){\cal X}^\nu(W)\rangle = - 2\alpha' \log(Z-W)
\label{calXcontract}
\end{equation}
The modified M{\"{o}}bius volume  indicates that one must fix two variables $\theta_i$ to be zero besides fixing three $z_i$. This is a consequence of the fact that the integrand of the previous expression is invariant under super-projective transformations. Actually recall that, for $\alpha' k_i^2 = \frac{1}{2}$, the theory is fully super-conformal invariant and this insures the elimination of all ghosts. 

Originally both the amplitudes in Eqs. (\ref{I1}) and (\ref{I10}) were supposed to describe mesons as the pions that are multiplets of $U(N_f)$ where $N_f$ is the number of flavours.  In order to include flavour, Chan and Paton~\cite{CP} proposed to multiply the previous amplitudes with the so-called Chan-Paton factors given by the  trace of the product of  the generators  of $U (N_f)$: $Tr ( T^{{a}_1} T^{{a}_2} \dots T^{{a}_N} )$. The total amplitude is then obtained by summing  over the $(N {-}1)!$ non-cyclic permutations of the external legs. Only a bit later the requirement of absence of ghosts forced the meson to become a tachyon and  the presence of a massless vector meson. At this point it was clear that the Chan-Paton factors in string theory describe colour rather than flavour. 

An early attempt to construct a $N$-pion amplitude with the Adler zeros was made in Ref. \cite{BrowerPLB71}, however, it was affected by the presence of tachyons as for the NS model.
On the other hand, keeping $k_i^2 \neq \frac{1}{2\ap}$, Eq. \eqref{I6} can be reformulated in the superconformal formalism in two dimensions as done for Eq. \eqref{I10}. The resulting $N$-point amplitude reads:
\begin{align}
A^{\rm NT-S}_N
&\,= \int \frac{\prod_{1=1}^N dZ_i}{dV_{abc}} \langle 0| \prod_{i=1}^N {\rm e}^{i k_i {\cal X}(z_i, \theta_i)} |0 \rangle \prod_{i=1}^N (Z_i - Z_{i+1}  )^{-\frac{1}{2} -\alpha' m_\pi^2 }
\nonumber \\
&\, = \int \frac{\prod_{1=1}^N d \theta_i  dz_i}{d V_{abc}} \prod_{i<j} (Z_i - Z_j )^{2\alpha' k_i k_j}\prod_{i=1}^N (Z_i - Z_{i+1}  )^{-\frac{1}{2} -\alpha' m_\pi^2 }
\label{I11}
\end{align}
This amplitude coincides with the models proposed by Neveu and Thorn \cite{NeveuThornPRL71} and by Schwarz \cite{SchwarzPRD72}, whence the super-script NT-S. The manifestly super-projective invariant expression (\ref{I11}) that we explicitly gave above was neither written down in \cite{NeveuThornPRL71, SchwarzPRD72} nor in \cite{FM}.

The integrand of this amplitude is super-projective invariant for any value of the mass $m_\pi$, but one often considers $m_\pi=0$ to start with as we will do in the following. 
The four-pion amplitude precisely reproduces the Lovelace-Shapiro model and the $N$-point amplitude generalizes it to  the scattering of $N$ pions. As before the total amplitude is obtained by multiplying it with a Chan-Paton factor and then by summing over the $(N{-}1)!$ non-cyclic permutations of the external legs. The vector meson, that would be massless in the NS model in $D=10$, becomes the massive $\rho$-meson. As a consequence and at variant from `modern' (super)string theory, the Chan-Paton factors describe flavour rather than colour degrees of freedom.

Unlike the amplitude in Eq. (\ref{I10}) that comes from a consistent string theory which is fully super-conformal invariant  and is free from ghosts, the one in Eq. (\ref{I11}), not being directly connected to any string theory, is only invariant under super-projective transformations and may exchange negative norm states (ghosts). However, as originally discussed in \cite{NeveuThornPRL71} and reproduced in a different form in Sect.  \ref{sect:NSKK}, four and six-point flavour ordered amplitudes in Eq. (\ref{I11}) can be derived  from those of the string amplitudes in Eq. (\ref{I10}) by suitably choosing the momenta of the tachyons along the six extra dimensions. Therefore we show that they are free of ghosts. Ghosts are on the other hand expected even in flavour-ordered amplitudes with a higher number of external legs and they may appear in the full/total six-pion amplitudes.

As shown in \cite{FairlieNPB72} and reviewed in App. \ref{explicit}, the integral over the variables $\theta_i$ can be performed and one gets:
\begin{equation}
A^{\rm NT-S}_N
= C_\textup{N-pt} \prod_{i=3}^{N-1} \int d z_i  \prod_{i=2}^N \prod_{i<j} (z_i -z_j)^{2\alpha' k_i k_j} \prod_{i=2}^{N-1} (z_i -z_{i+1})^{- \frac{1}{2}}  \sqrt{\det A}
\label{I12}
\end{equation}
where $A$ is the following antisymmetric matrix
\begin{eqnarray}
A = \left( \begin{array}{ccccc}  0  & \frac{2\alpha' K_2 K_3}{z_2-z_3} &   \frac{2\alpha' K_2 K_4}{z_2-z_4} & \dots &  \frac{2\alpha' K_2 K_{N-1}}{z_2-z_{N-1}} \\
- \frac{2\alpha' K_2 K_3}{z_2-z_3}  & 0 & \frac{2\alpha' K_3 K_4}{z_3-z_4} & \dots & \frac{2\alpha' K_3 K_{N-1}}{z_3-z_{N-1}} \\
- \frac{2\alpha' K_2 K_4}{z_2-z_4} & - \frac{2\alpha' K_3 K_4}{z_3-z_4} & 0 & \dots &  \frac{2\alpha' K_4 K_{N-1}}{z_4-z_{N-1}} \\
\dots & \dots & \dots & \dots & \dots \\
-     \frac{2\alpha' K_2 K_{N-1}}{z_2-z_{N-1}} & -\frac{2\alpha' K_3 K_{N-1}}{z_2-z_{N-1}} & 
-\frac{2\alpha' K_4 K_{N-1}}{z_4-z_{N-1}} & \dots & 0 \end{array} \right)    
\label{I13}
\end{eqnarray}
and the scalar product $K_i K_j$ is defined by
\begin{eqnarray}
&& K_i K_j = k_i k_j -\frac{1}{4\alpha'} ~~~~\text{if}~~~~j=i\pm 1 ~~;~~K_i K_j = k_i k_j ~~\text{if}~~~j \neq i\pm 1
\label{I14}
\end{eqnarray}
For an antisymmetric matrix,  such as $A$, $\sqrt{\det A}$ is also called the Pfaffian: ${\rm Pfaff}A$. The normalization factor in front of Eq. (\ref{I12}) is fixed by the factorization properties of the $N$-pion amplitude:
\begin{equation}
C_{\textup{N-pt}} = (\alpha')^{\frac{N-4}{2}} C_{\textup{4-pt}} ^{\frac{N}{2}-1} = (\alpha')^{-1} \left(-2\pi F_\pi^2 \right) ^{1- \frac{N}{2}}
\label{CN}
\end{equation}

The model enjoys several interesting properties such as the Adler zeroes and we show that it reduces to the non-linear $\sigma$-model in the field theory limit ($\alpha' \rightarrow 0$, keeping $F_\pi$ fixed)~\footnote{For a recent alternative  model that reproduces the non-linear $\sigma$-model in the field theory limit see Ref.~\cite{Carrasco:2016ldy}.}. It has also a low energy spectrum containing  the  mesons observed in hadronic processes, but some of the  masses and couplings deviate a bit from the observed ones. 

\begin{figure}[t]
\centering
~~\includegraphics[scale=0.3]{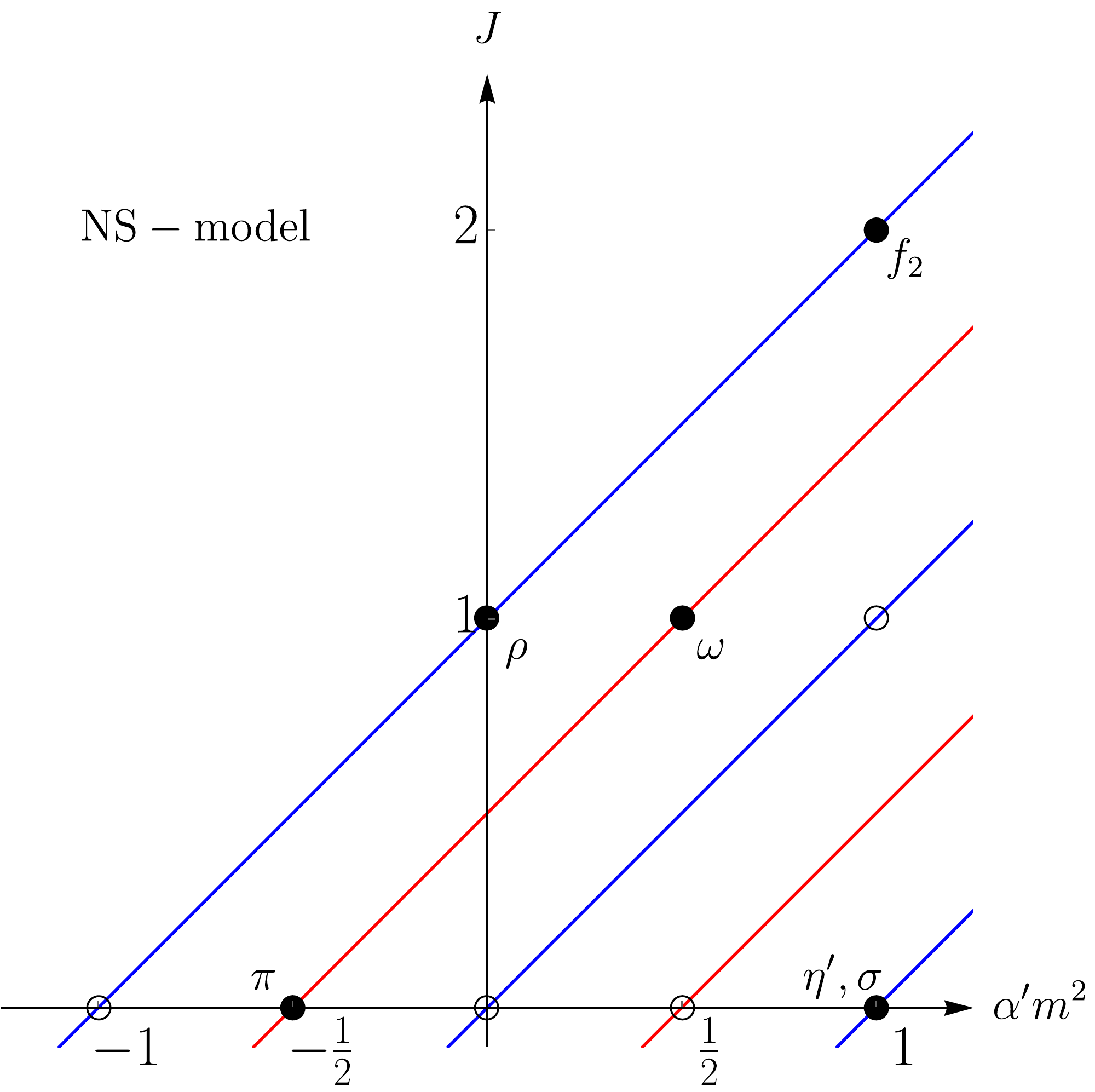}
~~~~~~~~
~~\includegraphics[scale=0.3]{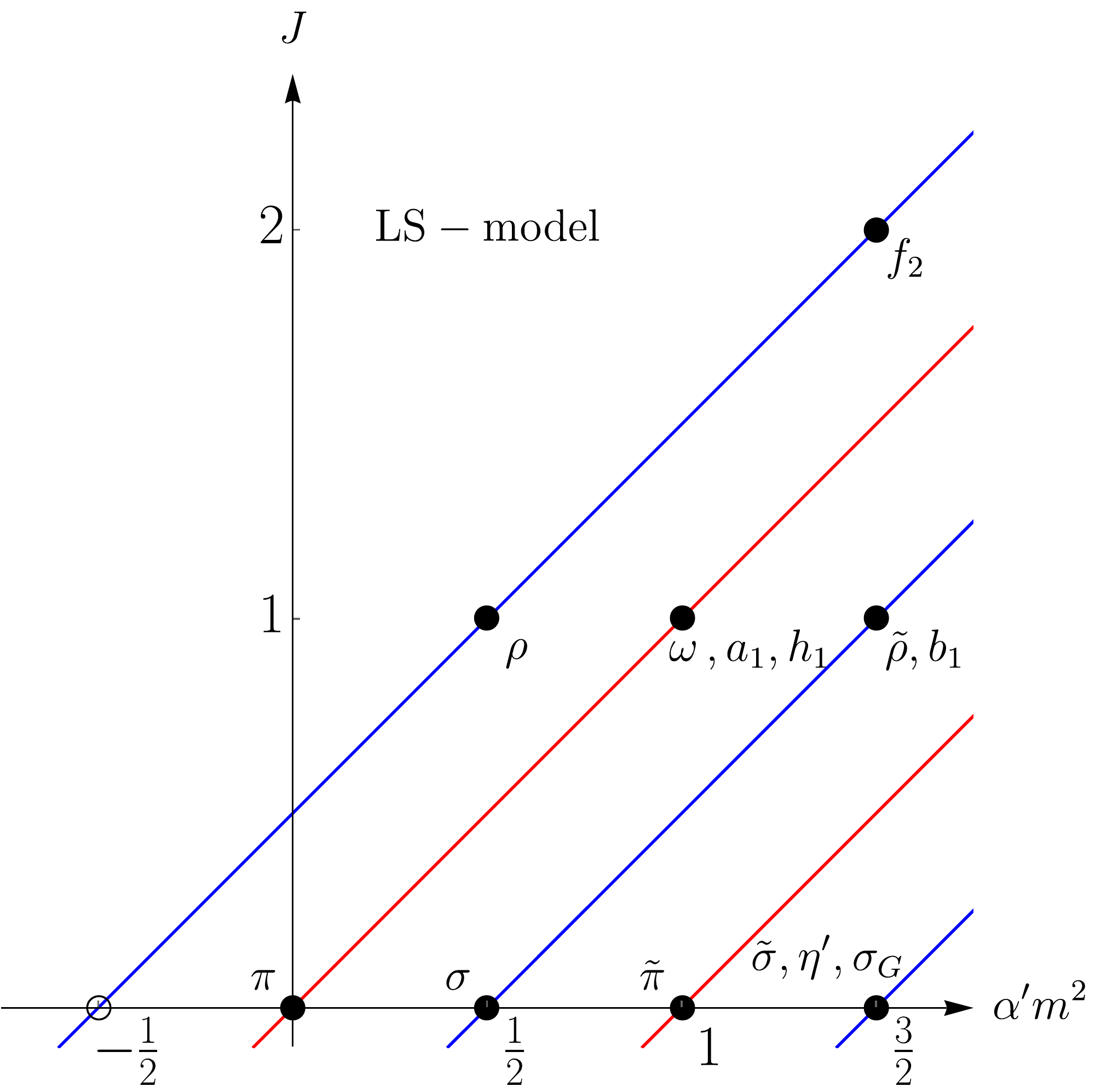}
\caption{Spectrum of NS (left) and LS (right) model in four dimensions, Regge trajectories in blue (red) have G-parity $+1$ ($-1$). Bullets represent `physical' states, open circles represent `missing' states.}
\label{spectra}
\end{figure}

As shown in the fig. (\ref{spectra}) on the right side, the model contains two types of Regge trajectories, one with integer intercept and another with half-integer one. They have the same slope that is equal to  $\alpha ' = {1}/{2m_\rho^2}$ for massless pions.  It incorporates a G-parity operator that is conserved. The particles lying on the Regge trajectories with half-integer intercept, as for instance the $\rho$ meson, have G-parity plus, while those lying on those with integer intercept, as the pion and the $\omega$, have G-parity minus. This implies that the amplitudes with an odd number of pions vanish as desired.

The particle with the lowest mass is the pion that we take to be massless. One can give a small mass to the pion, but then this would introduce ghosts already in the four-point amplitude. At the first excited level one has the $\rho$-meson with spin one and a scalar particle with spin zero that can be identified with a broad resonance that was called $\sigma$ in the '60's and  now is called $f^0$ with a mass around $500$ MeV. In the model that we here reconsider  it is instead degenerate with the $\rho$-meson that has a higher mass of $770$ MeV. The $\omega$-meson has spin 1 and lies on the pion trajectory. It has a mass higher than the $\rho$-meson, while in nature the two mesons are almost degenerate.  The spin $2$ on the $\rho$ Regge trajectory corresponds to the meson $f_2$ that has a mass of $1270$ MeV that is not so far from what the model under consideration predicts.

In Refs. \cite{FairlieNPB72,BrowerChuPRD73,SinghPasupathyPRD74} modifications to \eqref{I12} were considered in order to move the $\omega$ trajectory closer to the $\rho$ trajectory. Unfortunately this attempts failed to keep consistent factorizations and Adler zeroes.

Though perfectly consistent for low numbers of external legs $N$, the model is expected to be pathological and display negative norm (ghosts) states for $N$ large enough (e.g. $N\ge 8$).  In fact, the presence of the extra terms in the scattering amplitude and the value of the mass of the lowest state imply that the physical states are annihilated only by the lowest super-Virasoro generators $G_{1/2}$ and $L_1$. Yet for $N=4,6$, generalising \cite{NeveuThornPRL71}, we will show that `flavour-ordered' amplitudes are unitary in that they can be identified with amplitudes of would-be NS tachyons in $D=10$ with fixed/quantized internal (6-dimensional) momenta $q$, such that $2\alpha' q^2= 1$ and thus $m_4^2 = k^2=0$, with $k$ being the four-dimensional momentum.  

There are other features of the model we investigate that are in contradiction with QCD expectations. The model has linearly rising Regge trajectories for any value of the Mandelstam variable $s$, while in QCD, because of asymptotic freedom, one expects that they bend up for sufficiently negative values of $s$. See Refs. \cite{CKSZ} for a discussion of this point. The amplitudes (\ref{I12}) are too soft at large transverse momentum to describe hadrons. In order to cure this problem in Refs. \cite{AS,VYO} it has been proposed to sum over an infinite number of four-point amplitudes having Regge slopes  $\frac{\alpha'}{n}$. In principle,  this could also be 
done for the $N$-point amplitude, but, in this case, the procedure becomes  more laborious. Complementary approaches to hadron scattering have been recently proposed that rely on  holographic versions of QCD~\cite{Witten:1998zw,Sakai:2004cn,Bartolini:2016dbk,Aharony:2006da}, on open strings with massive ends \cite{Sonnenschein:2016pim, Sonnenschein:2017ylo, Sonnenschein:2018fph, Sonnenschein:2019bca} or on the S-matrix bootstrap \cite{Guerrieri:2019rwp}.


The paper is organised as follows.
In Sect. \ref{LS} we review the Lovelace-Shapiro model for the four-pion scattering amplitude and some of its properties, including its relation to the (non)linear $\sigma$ model at low energy.
In Sect. \ref{sect:NSKK}, following earlier work \cite{NeveuThornPRL71} we show that the  four- and six-point flavour-ordered amplitudes can be obtained from the correspondent amplitudes of the ten-dimensional NS model with a suitable choice of the momenta of the external tachyons in the six extra dimensions.
In Sect. \ref{SixPt} we consider in detail the six-pion amplitude \cite{NeveuThornPRL71} and we show that has Adler zeroes, correctly factorises in the product of two four-pion amplitudes at the pion pole in the three-pion channel  and in the field theory limit ($\alpha' \rightarrow 0$ with $F_\pi$ kept fixed) it reproduces the six-pion amplitude of the non-linear $\sigma$-model. We also present some three- and four-point amplitudes of excited states that can be obtained by factorisation in intermediate channels and we study tree-level unitarity of the $\p \s \to \p \s$ scattering amplitude.
In Sect. \ref{AdlerNpoint} we show that the $N$-pion amplitude (\ref{I12}) proposed in \cite{NeveuThornPRL71, SchwarzPRD72} has the Adler zeroes and the correct factorisation properties at the pion poles in various channels. In Sect. \ref{sect:unitarity} we extract from the eight-pion amplitude the scattering amplitude with four $\sigma$ particles and discuss its unitarity properties. Finally, in Sect. \ref{conclusions} we summarise our result and discuss possible connections with other approaches.
The paper contains also six Appendices. App.~\ref{A} is devoted to the super-projective transformations. In App.~\ref{explicit} we give some detail on the computation of the $N$-pion amplitude along the lines of \cite{NeveuThornPRL71, FairlieNPB72, SchwarzPRD72}.  In App.~\ref{group} we discuss the properties of the generators of $U(N_f)$ that enter in the Chan-Paton factors. App.~\ref{linear} is devoted to the linear and non-linear $\sigma$-model. In App.~\ref{field_theory_limit} we give a detailed derivation of the field theory limit of the six-pion amplitude.
Finally, in App.~\ref{ampl_appendix} we discuss some three- and four-point amplitudes extracted by factorisation from the six-pion amplitude.

\section{The Lovelace-Shapiro model}
\label{LS}

In this section we revisit the Lovelace-Shapiro model \cite{LOVELACE,SHAPIRO}. To this end we focus on the following `flavour'-ordered amplitude
\begin{equation}
A^{\text{(LS)}}[1,2,3,4]= C_\textup{4-pt} \frac{\G(1-\a_\r(s))\G(1-\a_\r(t))}{\G(1-\a_\r(s)-\a_\r(t))}
\label{eq:LS_general}
\end{equation}
where  $s =- (k_3+k_4)^2$, $t= - (k_2+k_3)^2$ and $\a_\r(s){\, = \,} \a_0{\, + \,} \ap s$ is the Regge trajectory of the vector meson $\r$, whose mass is given by $\ap m_\r^2{\,=\,}1{\,-\,}\a_0$. The full amplitude is obtained introducing the Chan-Paton factors
\begin{equation}
A^{\text{(LS)}}= 
\sum_{\s \in S_3} \Tr (T^{a_1} T^{a_{\s(2)}} T^{a_{\s(3)}} T^{a_{\s(4)}}) A^{\text{(LS)}}[1,\s(2),\s(3),\s(4)]
\label{eq:LS_general_full}
\end{equation}
In this paper we will consider a $U(N_f)$ flavour symmetry, although the isospin symmetry $SU(2)$ would be more appropriate to describe massless mesons.  The group-theoretical properties of the generators $T^a$ of $U(N_f)$ are summarised in App. \ref{group}.

One can require the presence of Adler zeroes, \ie that the amplitude vanishes when one of the momenta goes to zero. This condition can be obtained forcing the $\Gamma$ function in the denominator in \eqref{eq:LS_general} to be divergent, $ 2\a(m_\p^2){\,=\,}1 $, obtaining $\ap m_\p^2{\,=\,}1/2{\,-\,}\a_0$. This last equation has already been imposed in Eq. \eqref{I11}. In the case of $N=4$ we get
\begin{align}
A^{\text{(LS)}}[1,2,3,4] &{\,=\,} C_\textup{4-pt} \int_0^1 dz \, d\theta_2 \, d \theta_3 \left(1{\,-\,}z {\,-\,} \theta_2 \theta_3 \right)^{2\alpha' k_2 k_3-\alpha' m_\p^2 - \frac{1}{2}}
z^{2\alpha' k_3 k_4-\alpha' m_\p^2 - \frac{1}{2}} \nonumber \\
& {\,=\,} C_\textup{4-pt} \int_0^1 dz \left(1{\,-\,}z  \right)^{2\alpha' k_2 k_3 - \alpha' m_\p^2 - \frac{3}{2}} z^{2\alpha' k_3 k_4 - \alpha' m_\p^2 - \frac{1}{2}} \left(2\alpha' k_2 k_3 {-} \alpha' m_\p^2 {-} \frac{1}{2} \right) \nonumber \\
& {\,=\,} C_\textup{4-pt} \frac{\Gamma ( 1- \alpha_\rho (s)) \Gamma (1-\alpha_\rho (t))}{\Gamma (1-\alpha_\rho(s) - \alpha_\rho (t))}
\label{NP7}
\end{align}
where $ \alpha_\rho (s) =\a_0 +\alpha' s = \frac{1}{2} - \alpha' m^2_\p +\alpha' s$. Combining the Adler zero constraint with the $\r$ mass we get the following formulae for $\a_0$ and $\ap$ in terms of the masses
\begin{equation}
\a_0=\frac{m_\r^2-2 m_\p^2}{2(m_\r^2-m_\p^2)} \qquad \ap=\frac{1}{2(m_\r^2-m_\p^2)}
\label{a0_and_aprime}
\end{equation}
Constraints on $\a_0$ and the space-time dimensions $D$ can be obtained imposing that the residues at the poles be positive definite (tree-level unitarity). The residues of the full amplitude \eqref{eq:LS_general_full} at $\ap s = \ap s_n \equiv n{+}1{-}\a_0$ are the following
\begin{equation}
{\,-\,} \underset{s\,=\,s_n}{\Res} A^\textup{(LS)}=({-})^{n} \frac{C_\textup{4-pt}}{\ap n!} \left[ (T_{1234}{+}T_{1432}) \frac{\G(1{-}\a_\r(t))}{\G({-}n{-}\a_\r(t))}+
 (T_{1243}{+}T_{1342}) \frac{\G(1{-}\a_\r(u))}{\G({-}n{-}\a_\r(u))} \right]
\end{equation}
where $T_{ijkl}=\Tr (T_{a_i} T_{a_j} T_{a_k} T_{a_l})$. 
Writing $t$ and $u$ in terms of $s$ and the scattering angle $\q$:
\begin{eqnarray}
t= - \frac{s-4m_\pi^2}{2}(1-\cos \theta) ~~;~~u= - \frac{s-4m_\pi^2}{2}(1+\cos \theta)
\label{tutheta}
\end{eqnarray}
and using the relation: $\alpha' m_\pi^2 = \frac{1}{2} - \alpha_0$, the residue can be expanded in terms of Gegenbauer polynomials.\footnote{
Gegenbauer polynomials can be defined using the following recurrence relation
\begin{equation}
G_0^{(\a)}(x)=1 \quad , \quad
G_1^{(\a)}(x)=2 \a x \quad , \quad
n \, G_n^{(\a)}(x)=  2 x (n{+}\a{-}1) G_{n-1}^{(\a)}(x)-(n{+}2\a{-}2) G_{n-2}^{(\a)}(x)
\end{equation}
where $\a$ is related to the space-time dimensions by $\a=(D-3)/2$. 
Quite remarkably, in $D{\,=\,}5$, $G_n^{(1)}(\cos\q)=\c^\textup{SU(2)}_{j=n/2}(\q)$  with $\c^\textup{SU(2)}_{j}(\q)$ being the characters of $SU(2)$~\cite{probing}.} 
Unitarity requires the positivity of the coefficients of the expansion for amplitudes with the same initial and final state. Therefore we choose to identify the flavour indices as $a_1 {\,=\,} a_4 {\,=\,} a$ and $ a_2 {\,=\,} a_3 {\,=\,} b $. 
We consider the levels $n{\, = \,}0$ and $n{\, = \,}1$ obtaining the following residues
\begin{align}
{-}\underset{n=0}{\Res} \,A^\textup{(LS)}
= & \frac{C_\textup{4-pt}}{\ap}\left[
\frac{1-3 \alpha_0}{D-3} (f^{abc})^2\,G_1(x,D)+
(\a_0-1) (\widehat{d}^{abc})^2 \, G_0(x,D)\right] 
\label{eq:LS_res0} \\
{-}\underset{n=1}{\Res} \,A^\textup{(LS)}
= & \,
\frac{C_\textup{4-pt}}{\ap}
\Bigg[{-}\frac{9 \alpha _0^2}{(D{-}3) (D{-}1)} (\widehat{d}^{abc})^2 G_2(x,D)
{+}\frac{3 \a_0(\a_0{-}1) }{D-3}(f^{abc})^2 G_1(x,D){+} \nonumber \\
&{+}
\frac{ \a_0 (2 D{-}2{-}8\a_0 {-}D \a_0)}{2 (D{-}1)}(\widehat{d}^{abc})^2G_0(x,D)\Bigg]
\qquad (a,b\text{ no sum})
\label{eq:LS_res1}
\end{align}
where $x{\,=\,}\cos \q$, $G_n(x,D)$ are Gegenbauer polynomials in $D$ space-time dimensions, $f^{abc}$ are the structure constants of $U(N_f)$, $\widehat{d}^{abc}$\footnote{We have chosen the normalisation $\Tr (T^a T^b)=\d^{ab}$ where $a$ and $b=0,1, \dots N_f^2{\,-\,}1$. In our definition of $f^{abc}= i \sqrt{2} \Tr ([ T^a, T^b] T^c)$, $\widehat{d}^{abc}= \sqrt{2} \Tr (\{ T^a, T^b\} T^c)$, $T^0= \idnty /\sqrt{N_f}$, $\widehat{d}^{abc}=d^{abc}$ if all the indices represent $SU(N_f)$  generators while $\widehat{d}^{0ab}= \sqrt{2/N_f} \d^{ab}$.} is the symmetric invariant tensor of $U(N_f)$ and the flavour indices $a$, $b$ are not summed over, while there is a sum over the flavour index $c$. Considering only states with even spin for now, we get the constraints
\begin{equation}
C_\textup{4-pt}\leq 0 \quad, \quad \a_0 \leq 1 \quad, \quad  D \leq D_c=2\frac{1+4 \a_0}{2-\a_0}
\label{eq:min_conditions}
\end{equation}
The possible values of $\a_0$ with an integer number of dimensions compatible with the bounds above are
\begin{gather}
\a_0(D{=}4)=\frac{1}{2} \qquad
\a_0(D{=}5)=\frac{8}{13} \qquad
\a_0(D{=}6)=\frac{5}{7} \qquad
\a_0(D{=}7)=\frac{4}{5} \\
\a_0(D{=}8)=\frac{7}{8} \qquad
\a_0(D{=}9)=\frac{16}{17} \qquad
\a_0(D{=}10)=1
\label{eq:possible_a0}
\end{gather}
$D=3$ is excluded from our analysis, since $\alpha = (D-3)/2=0$ and  the Gegenbauer polynomials are not defined (except for $G_0=1$). Notice that the choice $D{\, = \,}10$ reproduces the four-tachyon amplitude in the NS model.

As far as we know, barring its secret relation to the NS model in $D=10$ to be unveiled later on, there is no analytic proof of the positivity of the coefficients at an arbitrary level. We have considered the levels up to $n{\, = \,}20$ and we found that the conditions \eqref{eq:min_conditions} are enough to get positivity of the coefficients for all choices of $(\a_0,D_c)$ shown in \eqref{eq:possible_a0}. 

For values of $\a_0$ greater than $1/2$ the pion is a tachyon, while for $\a_0$ less than $1/2$ the critical dimension becomes smaller that four, therefore, $\a_0 = 1/2$ is a very special choice:
\begin{equation}
\a_0=\frac{1}{2} \quad , \quad D_c=4
\qquad \Longrightarrow\qquad  
m_\p^2=0 \quad , \quad m_\r^2=\frac{1}{2\ap}
\end{equation}
Using Eq \eqref{a0_and_aprime} and the experimental values of the pion and $\r$ masses, the value of $\a_0$ obtained is close to $1/2$. For this reason, we fix the intercept $\a_0$ of the $\r$ trajectory to be $1/2$ for the rest of the paper.

The residue at a general pole gives us informations on the quantum numbers of the states lying on the $\r$ and daughters trajectories
\begin{align}
-\underset{s=s_n}{\Res} \, A^\textup{(LS)}
& =  \frac{C_\textup{4-pt}}{\ap n!} \left[ 
(\widehat{d}^{abc})^2( F_n(x){+} F_n({-}x) )
+
(f^{abc})^2( F_n(x){-} F_n({-}x) )
\right]
\end{align}
where $a$ and $b$ are again not summed and $F_n(x)$ is a polynomial of degree $n{\,+\,}1$
\begin{equation}
\ap s_n=n+\frac{1}{2} \quad , \quad
F_n (x)=(-)^{n} \frac{\G\left(\frac{2n{+}3}{4}-x\frac{2n{+}1}{4}\right)}{\G\left(-\frac{2n{+}1}{4}-x\frac{2n{+}1}{4}\right)} 
\end{equation}
The terms $F_n(x){\pm} F_n({-}x)$ can be expressed as linear combinations of even (${+}$) or odd (${-}$) Gegenbauer polynomials. Therefore, intermediate particles with even (odd) spin always couple with $\widehat{d}$ ($f$). In the case of an $U(2)$ flavour group they have isospin $I{\, = \, }0$ (${\, I= \, }1$).

The residue in Eq. \eqref{eq:LS_res0} shows the presence of two states, the vector $\r$ and a scalar. In the case of $U(2)$  they have isospin $I{\, = \, }0$ and $I{\, = \, }1$ respectively. Therefore we identify the scalar with the $\s$ meson.

In the next sections we will factorise pairs of pions to construct amplitudes with $\r$ and $\s$ as external states. To this end we compute here the three-point amplitudes involving two pions and one $\r$ or $\s$ from the LS amplitude via factorisation and obtain
\begin{equation}
A_{\p \p \s}= - \h_\s \sqrt{\frac{-C_\textup{4-pt}}{2\ap}}\, \widehat{d}^{a_1 a_2 a_3}
\quad , \quad
A_{\p \p \r}= - \h_\r  \sqrt{-C_\textup{4-pt}} \e_3{\cdot}(k_1-k_2) \, f^{a_1 a_2 a_3}
\label{eq:APiPiRho_APiPiSigma}
\end{equation}
where $\h_\s$ and $\h_\r$ are undetermined signs.

\subsection{Comparison with linear and non-linear \texorpdfstring{$\sigma$}{Sigma}-model}
\label{LS_sigma_models}
The limit $\ap {\,\to\,} 0$ of the amplitude \eqref{eq:LS_general} is consistent for $\a_0{\,=\,}1$ or $1/2$. The first case correspond to the limit $m_\p^2 \to - \infty$, while the second to $m_\r^2 \to +\infty$. For all the intermediate values of $\a_0$ between $1$ and $1/2$, $m_\p^2 \to - \infty$ and $m_\r^2 \to +\infty$ simultaneously. 
For $\a_0{\,=\,}1/2$ the limit $\ap {\, \to \,}0$ gives us the following amplitude
\begin{equation}
A^{\text{(LS)}}[1,2,3,4]\xrightarrow{\ap \to 0}-\ap \pi \, C_\textup{4-pt} (s+t) +\dots
\label{eq:LS_NLSM}
\end{equation}
This limit matches the correspondent amplitude in the non-linear $\s$-model \eqref{eq:NLSM_4pt_color_ordered} if we set $C_\textup{4-pt}$ to be
\begin{equation}
C_\textup{4-pt}=-\frac{1}{2 \p \ap F_\p^2}
\label{eq:C4_NLSM}
\end{equation}
We now consider if there is agreement between the linear $\s$-model and the LS amplitude with flavour symmetry $SU(2)$.
First of all, the LS amplitude can be written in such a way to expose separately the poles at $m_\r^2 =m_\s^2 = \frac{1}{2\ap}$ in the $s$- and $t$-channel:
\begin{equation}
A^{\text{(LS)}}[1,2,3,4]= 
\frac{s+t}{2 \p F_\p^2}
\left(\frac{1}{\frac{1}{2}-\ap s} +\frac{1}{\frac{1}{2}-\ap t}\right)
\frac{\G(\frac{3}{2}-\ap s)\G(\frac{3}{2}-\ap t)}{\G(2-\ap s-\ap t)}
\end{equation}
Then, since the ratio of the $\Gamma$ functions is equal to $\frac{\pi}{4}$ in the limit for $\alpha' \rightarrow 0$, one obtains
\begin{equation}
A^{\text{(LS)}}[1,2,3,4]\sim
-\frac{m_\s^2}{4 F_\p^2}
\left(2+\frac{m_\s^2/2}{s-m_\s^2}+\frac{m_\s^2/2}{t-m_\s^2}+\frac{t+m_\r^2/2}{s-m_\r^2}+\frac{s+m_\r^2/2}{t-m_\r^2}\right)
\label{eq:LS_sigma_model}
\end{equation}
where we have used the relations  $m_\r^2 =m_\s^2 = \frac{1}{2\ap}$  to eliminate the dependence on $\alpha'$. Notice that in the limit $m_\r{\, = \,}m_\s{\, \to\,} \infty$ we obtain again the non-linear $\s$-model. The amplitude \eqref{eq:LS_sigma_model} has
the Adler zero only if we keep both $\s$ and $\r$. Therefore it cannot be compared with the linear $\s$-model in \eqref{eq:Apppp_CO_LSM} but rather to an extended version which also includes the $\r$ meson.

\section{The Neveu-Schwarz model with massless pions}
\label{sect:NSKK}

In this section, following a slightly different approach from earlier suggestion \cite{NeveuThornPRL71}, we would like to show that the flavour-ordered LS amplitude (\ref{eq:LS_general}) can be identified with an amplitude of would-be NS tachyons in $D{\,=\,}10$ with fixed/quantized internal (six-dimensional) momenta $q$, such that $2\alpha' q^2{\,=\,} 1$ and thus $m_4^2 {\,=\,} k^2{\,=\,}0$, with $k$ being the  four-dimensional momentum. In this approach pions should be regarded as KK modes of the NS tachyon. The crucial property that follows from this approach is unitarity, that is guaranteed by the  unitarity of the NS string in $D{\,=\,}10$.

After reproducing the four-point LS amplitude (\ref{eq:LS_general}) in this approach, we will address higher-point amplitudes \cite{NeveuThornPRL71, FairlieNPB72, SchwarzPRD72, SinghPasupathyPRD74}. In particular we will explicitly reconstruct the six-point `flavour-ordered' amplitude \cite{BrowerChuPRD73} and again show that inherits unitarity from the parent NS string. Since at most six linearly independent $q$'s are available, beyond  six-points unitarity is not guaranteed anymore\footnote{In the phenomenological proposals \cite{NeveuThornPRL71, FairlieNPB72, SchwarzPRD72, SinghPasupathyPRD74} this point was largely overlooked.}. To have more $q$'s at disposal, one can formally generalise the KK approach and consider a NS string in more than ten dimensions. Since in this case the NS model has ghosts we expect the appearance of ghosts in flavour-ordered amplitudes with $N\geq 8$ pions.

In order to show the relation between NS tachyons and LS pions, let us consider the NS tachyon vertex operator in ten dimensions, {\it viz.}
\begin{equation}
V^{(0)}_\pi =\, C_\p\, P{\cdot}\Psi \,e^{iPX} \quad , \quad
V^{(-1)}_\pi =\, C_\p \,e^{-\varphi} \,e^{iPX}
\end{equation}
where the super-scripts $(0)$ and $(-1)$ denote the super ghost `picture' and $C_\p$ is a normalisation constant to be fixed later on. 
If one decomposes the tachyonic 10-dimensional momentum as $P = (k, q)$, with $k^2=0$ in four non compact dimensions and $q^2 = 1/2\alpha'$ for the six internal dimensions, one can imagine $q$ playing the role of Kaluza-Klein momentum. We also decompose the (super)coordinates as 
$\Psi = (\psi_{_X}, \psi_{_Y})$ where $X$ denote the four-dimensional bosonic non-compact coordinates and $Y$ the six-dimensional internal ones.
More explicitly 
\begin{equation}
\label{LSpionKKtach}
V^{(0)}_\pi = C_\p (k{\cdot}\psi_{_X} + q{\cdot}\psi_{_Y}) e^{i(kX+qY)} \,\quad , \quad
V^{(-1)}_\pi = C_\p \, e^{-\varphi}\, e^{i(kX+qY)}
\end{equation}
The flavour-ordered LS amplitude can be reproduced from the correspondent NS amplitude by taking the six-dimensional part of the momenta of the four external pions to satisfy the relation $2\alpha' q_i^2= 1$ together with the conditions:
\begin{equation}
\label{qconditions} 
2\alpha' q_i{\cdot}q_{i\pm 1} = -\frac{1}{2} \qquad \text{ while } \qquad
q_i{\cdot}q_j = 0 \text{ if } j\neq i\pm 1
\end{equation}
For the flavour-ordered four-point amplitude, a solution that satisfies the above constraints is the following
\begin{align}
q_1&= \frac{1}{2\sqrt{\alpha'}}(1 ,-1, 0, 0,0,0) \qquad  
q_2 = \frac{1}{2\sqrt{\alpha'}}(0 , 1,-1, 0,0,0) \\
q_3&= \frac{1}{2\sqrt{\alpha'}}(0 , 0, 1,-1,0,0) \qquad 
q_4 = \frac{1}{2\sqrt{\alpha'}}(-1, 0, 0, 1,0,0)
\end{align}
This solution is unique up to $O(6)$ transformations\footnote{Different but equivalent choices can be made as in \cite{NeveuThornPRL71, FairlieNPB72, SchwarzPRD72, SinghPasupathyPRD74}.}.

For the flavour-ordered six-point amplitude one can make a similar choice for the first three $q$'s: $q_1'=q_1, q_2'=q_2, q_3'=q_3$ and then take
\begin{equation}
q_4' = \frac{1}{2\sqrt{\alpha'}} ( 0, 0, 0, 1,-1, 0)\quad 
q_5' = \frac{1}{2\sqrt{\alpha'}} ( 0, 0, 0, 0, 1,-1)\quad
q_6' = \frac{1}{2\sqrt{\alpha'}} (-1, 0, 0, 0, 0, 1)
\end{equation}
Again the solution is unique up to $O(6)$ transformations. There is no way to find a solution to (\ref{qconditions}) for more than six external legs.
One can easily verify that both the flavour-ordered four-point and six-point amplitudes are independent of the `pictures' assigned to the various vertices and the flavour-ordered six-point amplitude in \eqref{A6igig} is neatly reproduced.

As a consequence of the underlying origin, the resulting flavour-ordered amplitudes are `unitary' in $D\leq 10$ like their parent NS amplitudes.
For the four-point amplitude the internal momenta span four directions, so one may expect unitarity for $k_i^2 =0$ up to $D=6 = 10 - 4$. By the same token, in the case of the six-point amplitude six internal directions are necessary to make things work, so one may expect unitarity for $D\leq 4$. Pushing this argument further to eight points, an amplitude built following the above procedure would be unitary for $D\leq 2$.  

In Ref.~\cite{HT}  a different inequivalent choice of the internal momenta is made, \ie $q_{2i} = q$ and $q_{2i+1} = -q$ with $2\alpha' q^2{\,=\,} 1$, in order for the resulting `pions' to be massless. The obvious advantage of this approach is that it immediately generalises to $N$-point amplitudes with $N$ even and gives zero for $N$ odd. The fatal drawback is the absence of Adler zeros that prevents the identification of the `pions' with (quasi) Goldstone bosons. In particular, already at the level of the four-points the amplitude proposed in \cite{HT} does \textit{not} coincide with the LS amplitude.

\subsection{Four-point amplitude}
Let us show explicitly how the LS amplitude can be reproduced by computing the following flavour-ordered amplitude
\begin{align}
A^{(1234)}_{4\pi}=&\,C_D \,C_\p^4 \int_0^1 dx \, \langle V^{(-1)}_\pi(\infty)\,V^{(-1)}_\pi(1) \,V^{(0)}_\pi(x)\,V^{(0)}_\pi(0)\>= \nonumber\\
=&\,C_D \,C_\p^4\, (2 \ap P_3 {\cdot} P_4)\, \int_0^1 dx\, x^{2 \alpha' P_3{\cdot}P_4-1} (1-x)^{2 \alpha' P_2{\cdot}P_3}
\end{align}
where  $C_D$ is the disk normalisation constant and we have used the correlators that follow from Eq.~\eqref{calXcontract} to compute the contractions with $X$ and $\Psi$. 
The superscript $1234$ encodes the ordering of the pions up to cyclic permutations and inversions. Integrating over $x$ we get
\begin{equation}
A^{(1234)}_{4\pi}=\,C_D \,C_\p^4\, \frac{\Gamma(-\alpha' S) \Gamma(-\alpha' T)}{\Gamma(1-\alpha' S-\alpha' T)}=\,C_D \,C_\p^4\, \frac{\Gamma(\frac{1}{2}-\alpha' s) \Gamma(\frac{1}{2}-\alpha' t)}{\Gamma(-\alpha' s-\alpha' t)}
\label{fourpoint}
\end{equation}
where $S$, $T$ and $s$ , $t$ are respectively the ten- and four-dimensional Mandelstam invariants. The amplitude in Eq. \eqref{fourpoint} reproduces the LS amplitude in  Eq. \eqref{NP7} with the identification $C_\textup{4-pt}= \,C_D \,C_\p^4$. Similarly for an $N$-pion amplitude one obtains $C_\textup{N-pt}= \,C_D \,C_\p^N$. Using Eq. \eqref{CN} we can fix the disk and the vertex operator normalisations to be
\begin{equation}
C_D=\frac{1}{\ap^2 C_\textup{4-pt}}=-\frac{2\p F_\p^2}{\ap} \quad , \quad
C_\p= \sqrt{\ap C_\textup{4-pt}} =  (- 2\p F_\p^2)^{-1/2}
\end{equation}
In order to get the complete four-point amplitude, one has to multiply each flavour-ordered amplitude by the corresponding Chan-Paton factor and sum over the six non-cyclic permutations, 
\begin{equation}
A^{tot}_{4\pi} = A^{(1234)}_{4\pi} (T_{1234} + T_{4321}) + A^{(1342)}_{4\pi} (T_{1342} + T_{3124}) + A^{(1423)}_{4\pi} (T_{1423}+ T_{4132})
\end{equation} 
where $T_{1234} = \Tr(T_1T_2T_3T_4)$ and similarly for the rest. Notice that $A^{(1234)}_{4\pi} = A^{LS}(s,t)$ then $A^{(1342)}_{4\pi} = A^{LS}(u,s)$ and $A^{(1423)}_{4\pi} = A^{LS}(t,u)$. One should, however, keep  in mind that each flavour-ordering corresponds to a different assignment of internal momenta $q$'s satisfying Eqs.$\,$\eqref{qconditions}. From the 10-dimensional perspective this means that internal components are reshuffled keeping the same space-time components. For instance $A^{(1342)}_{4\pi} = A^{LS}(u,s)$ requires $P_1 = (k_1, q_1)$, $P_2=(k_2, q_4)$, $P_3 = (k_3, q_2)$, and $P_4=(k_4, q_3)$. Similar reshuffling is also necessary  for $A^{(1423)}_{4\pi} = A^{LS}(t,u)$, an issue that was not addressed earlier \cite{NeveuThornPRL71, FairlieNPB72}.

Postponing the detailed derivation of the (unitary) flavour-ordered six-point amplitude to the next Section, we would like to briefly comment on how flavour-ordered $N$-point amplitudes with $N\ge 8$ can be formally obtained in the KK approach. The idea is to continue to impose the conditions $2\alpha' q_i^2= 1$, in order to have massless pions, and $2\alpha' q_i{\cdot}q_{i\pm 1} = -\frac{1}{2}$ as well as $q_i{\cdot}q_j = 0$ if $j\neq i\pm 1$ in order to have Adler zeros. This would be impossible in $D=10$ and would require a NS string in 
at least $D=4+N$ for an $N$-point amplitude where the theory is known to be non-unitary and expose `ghosts'. A possible way out is to consider non-critical NS strings, that would require the coupling  to the ${\cal N}=1$ super Liouville fields.  Another one is to replace the six `internal' dimensions with some interacting CFT that include more than six independent operators of conformal dimension $h = 1/2$ satisfying the desired algebra to `dress' the NS tachyon vertex and make it massless in $D=4$.

\section{The six-pion amplitude: factorisations and low energy limit}
\label{SixPt}

In this section we consider the six-pion amplitude and we discuss its properties that extend those already discussed in the case of the four-pion amplitude.  In particular, as already observed in \cite{NeveuThornPRL71, SchwarzPRD72} using a different procedure, we show that it has Adler zeroes and factorises into the product of two four-pion amplitudes in the three-particle channels. Moreover we show that it reduces to the six-point amplitude of the non-linear $\sigma$-model in the field theory limit. Furthermore, by factorising it in two-pion channels, we construct four-point amplitudes involving the states $\s$ and $\r$. 

In general a six-point amplitude is given by the sum of 120 terms involving products of the Chan-Paton factors with the flavour stripped amplitudes
\begin{equation}
A_{6\p}=\sum_{\s \in S_5} A_{6\p}^{(1 \s(23456))} \Tr[T_1 T_{\s(2)}\dots T_{\s(6)}]
\end{equation}
Let us consider here the fundamental permutation $(123456)$. Since the integrand of the six-pion amplitude is invariant under super-projective transformations, we can choose any pair of $\theta_i$ to vanish. If we choose $\theta_1 =\theta_2 =0$, together with $z_1 = \infty, z_2=1$ and $z_N=0$, and we integrate over the remaining Grassmann variables, we obtain three terms
\begin{eqnarray}
A_{6\p}^{(123456)}= && \, C_\textup{6-pt}  \int_0^1 dz_3 \int_0^{z_3} dz_4 \int_0^{z_4} dz_5 \,
(1- z_3)^{-\ap s_{23}- \frac{1}{2}} (1-z_4)^{-\ap s_{24}}
(1-z_5)^{-\ap s_{25}}  \nonumber \\
&& \times 
(z_3 - z_4 )^{-\ap s_{34} - \frac{1}{2}}
(z_3 - z_5 )^{-\ap s_{35}} 
 z_3^{-\ap s_{36}} 
(z_4 - z_5 )^{-\ap s_{45}- \frac{1}{2}}
z_4^{-\ap s_{46}}
z_5^{-\ap s_{56}- \frac{1}{2}}
\nonumber \\
&& \times \left[
\frac{ \left(\ap s_{34} + \frac{1}{2}   \right) \left(\ap s_{56} + \frac{1}{2}   \right)}{(z_3-z_4)z_5} - \frac{\ap^2 s_{35} s_{46}}{(z_3-z_5) z_4 }
+ \frac{ \ap s_{36}  \left(\ap s_{45} + \frac{1}{2}   \right) }{ 
z_3 (z_4 -z_5)} \right]
\label{NS11a}
\end{eqnarray}
$C_\textup{6-pt}$ is a constant that is determined below. Changing variables to $z_3 {\,=\,} \a$, $z_4 {\,=\,} \a \, \b$ and $z_5 {\,=\,} \a\, \b \,\g$ we get
\begin{align}
A_{6\p}^{(123456)}=&\, C_\textup{6-pt} 
\int_0^1 d\alpha\, \int_0^1 d\beta \, \int_0^1 d\gamma\,
\alpha^{-\alpha' s_{12} - \frac{1}{2}}
(1-\alpha)^{-\alpha' s_{23} - \frac{1}{2}}
\beta^{-\alpha' s_{123} }
(1-\beta)^{ - \alpha' s_{34} - \frac{1}{2}} 
\nonumber \\
& \times \gamma^{ -\alpha' s_{56} - \frac{1}{2}}
(1-\gamma)^{-\alpha' s_{45} - \frac{1}{2} }
(1- \alpha \beta)^{-\alpha' s_{24}}
(1-\beta \gamma)^{-\alpha' s_{35}} 
(1-\alpha \beta \gamma)^{ -\alpha' s_{25} } 
\nonumber \\
& \times  
\left[
\frac{ \left(\alpha' s_{34}+\frac{1}{2} \right)\left( \alpha' s_{56}+\frac{1}{2} \right) }{\alpha \beta \gamma (1-\beta)}
-\frac{ \alpha'^2 s_{35} s_{46} }{\alpha \beta  (1-\beta \gamma)} + \frac{\alpha's_{36} \left(\alpha's_{45}+\frac{1}{2} \right)}{\alpha \beta  (1-\gamma)} 
\right]
\label{eq:paolo_6pt}
\end{align}
This expression seems to have a pole $\alpha' s_{12} = -{1}/{2}$ that corresponds to a tachyon. The appearent poles at $ s_{34}, s_{45}, s_{56}= -{1}/{2 \alpha'}$ in the first and third terms, exposed by integrating over the variables $1{-}\beta$, $1{-}\gamma$, and $\gamma$, are cancelled by the corresponding numerators.

A way to show the absence of the pole at $\alpha' s_{12} {\,=\,} -{1}/{2}$, present in all three terms in \eqref{eq:paolo_6pt} and exposed by integrating over $\alpha$, is to consider the choice $\theta_1 {\,=\,} \theta_6{\,=\,}0$. Using the variables $\a$, $\b$ and $\g$ as in Eq. \eqref{eq:paolo_6pt} we get
\begin{align}
A_{6\p}^{(123456)}=&\, C_\textup{6-pt} 
\int_0^1 d\alpha\, d\beta \, d\gamma\,
\alpha^{-\alpha' s_{12} - \frac{1}{2}}
(1-\alpha)^{-\alpha' s_{23} - \frac{1}{2}}
\beta^{-\alpha' s_{123} }
(1-\beta)^{ - \alpha' s_{34} - \frac{1}{2}} 
\nonumber \\
& \times \gamma^{ -\alpha' s_{56} - \frac{1}{2}}
(1-\gamma)^{-\alpha' s_{45} - \frac{1}{2} }
(1- \alpha \beta)^{-\alpha' s_{24}}
(1-\beta \gamma)^{-\alpha' s_{35}} 
(1-\alpha \beta \gamma)^{ -\alpha' s_{25} } 
\nonumber \\
& \times  \left[  \frac{  \left( \ap s_{23}+\frac{1}{2}  \right) \left( \ap s_{45} +\frac{1}{2} \right)  }{(1-\a)\b (1-\g)}
-\frac{\ap^2 s_{24} s_{35}  }{(1- \a \b ) (1 - \b \g)} +
\frac{\ap s_{25}\left(\ap s_{34} +\frac{1}{2} \right)}{(1-\a \b \g) (1-\b)} 
\right]
\label{A6igig}
\end{align}
This expression of the amplitude does not explicitly show the presence of a tachyon for $\alpha' s_{12} = -{1}/{2}$. Therefore, this pole should also be absent
in \eqref{eq:paolo_6pt}.

\subsection{Three-particles channel, Adler zero and low energy limit}

As mentioned above, the flavour-ordered amplitude \eqref{eq:paolo_6pt} factorizes, near $s_{123}{\,=\,}0$, into the product of two LS amplitudes. This can be seen by integrating the amplitude around $\b {\,\to\,} 0$ for $s_{123} \sim 0$ and obtaining
\begin{equation}
A_{6\p}^{(123456)}\xrightarrow{s_{123}\to 0} -\frac{C_\textup{6-pt}}{\ap} 
\frac{\G(\frac{1}{2}-\ap s_{12})\G(\frac{1}{2}-\ap s_{23})}{\G(\ap s_{13})}\frac{1}{s_{123}}
\frac{\G(\frac{1}{2}-\ap s_{45})\G(\frac{1}{2}-\ap s_{56})}{\G(\ap s_{46})}
\end{equation}
The factorization provides also a way to fix the constant $C_\textup{6-pt}$
\begin{equation}
C_\textup{6-pt}= \ap C_\textup{4-pt}^2 = \frac{1}{4\p^2 \ap F_\p^4}
\label{eq:C6asC4}
\end{equation}
Another important feature of the six-pion amplitude is the presence of Adler zeroes. This can be shown by letting one of the external momenta vanish. Anyone would be enough since invariance of the amplitude under cyclic transformation ensures that the result holds for any external leg. Let us consider \eqref{A6igig} in the limit $k_2\to 0$, keeping $s_{12}$ and $s_{23}$ non-vanishing for now. We get
\begin{align}
A_{6\p}^{(123456)}=&\, C_\textup{6-pt} \!
\int_0^1 \! d\beta \, d\gamma\,
\gamma^{ -\alpha' s_{56} - \frac{1}{2}}
\beta^{-\alpha' s_{13} -1}
(1{-}\beta)^{ - \alpha' s_{34} - \frac{1}{2}} 
(1{-}\gamma)^{-\alpha' s_{45} - \frac{3}{2} }
(1{-}\beta \gamma)^{-\alpha' s_{35}} 
\nonumber \\
& \times \int_0^1 d\alpha \,
\alpha^{-\alpha' s_{12} - \frac{1}{2}}
(1-\alpha)^{-\alpha' s_{23} - \frac{3}{2}}
 \left( \ap s_{23}+\frac{1}{2}  \right) \left( \ap s_{45} +\frac{1}{2} \right)
\end{align}
Integrating over $\a$ we get
\begin{align}
A_{6\p}^{(123456)}=&\, C_\textup{6-pt} 
\int_0^1  d\beta \, d\gamma\,
\gamma^{ -\alpha' s_{56} - \frac{1}{2}}
\beta^{-\alpha' s_{13} -1}
(1{-}\beta)^{ - \alpha' s_{34} - \frac{1}{2}} 
(1{-}\gamma)^{-\alpha' s_{45} - \frac{3}{2} }
(1{-}\beta \gamma)^{-\alpha' s_{35}} 
\nonumber \\
& \times
\frac{\G(\frac{1}{2}-\alpha' s_{12}) \G(\frac{1}{2}-\alpha' s_{23})}{\G(-\alpha' (s_{12}+ s_{23}))}
\left( \ap s_{45} +\frac{1}{2} \right)
\end{align}
When $k_2$ vanishes the $\Gamma$ function in the denominator is divergent and therefore, the whole amplitude vanishes. The same computation can be repeated with the momentum $k_5$. To find explicitly the Adler zeroes for the other momenta we need to use different parametrizations of the amplitude. The Adler zeroes when $k_3$ or $k_6$ vanish can be shown considering Eq. \eqref{eq:paolo_6pt_X2X6Y2} while for $k_4$ using Eq. \eqref{eq:paolo_6pt_X1X3Y3}.

The last feature we will study is the field theory limit that was not considered in \cite{NeveuThornPRL71, FairlieNPB72, SchwarzPRD72,BrowerChuPRD73, SinghPasupathyPRD74}. The computation is quite laborious and the details are given in appendix \ref{field_theory_limit}. To arrive at the final result, one has to  
rewrite the amplitude \eqref{eq:paolo_6pt} using variables associated to the planar channels and manipulating the amplitude in order to separate the massless poles. After this manipulation one can perform the field theory limit without the presence of ambiguous divergences. The result is the following
\begin{align}
A_{6\p}^{(123456)}=&-\frac{1}{4 F_\p^4} \Bigg[\frac{(s_{12}{+}s_{23})(s_{34}{+}s_{45})}{s_{123}}{+}\frac{(s_{23}{+}s_{34})(s_{45}{+}s_{56})}{s_{234}}{+}\frac{(s_{34}{+}s_{45})(s_{56}{+}s_{16})}{s_{345}}{+}\nonumber\\
& {-} (s_{12}{+}s_{23}{+}s_{34}{+}s_{45}{+}s_{56}{+}s_{16})\Big]+\calO(\ap) 
\label{eq:low_energy_limit}
\end{align}
which reproduces the correspondent amplitude in the non-linear $\s$-model \eqref{eq:6ptNLSM} with the exact overall coefficient. This result is expected thanks to the invariance of the amplitude under cyclic transformations, the factorisation of the six point amplitude in two LS amplitudes (which ensures the presence of the first three terms in \eqref{eq:low_energy_limit} with the right coefficients) and the Adler zeroes (which force the linear terms in $s_{ij}$ to have the coefficients shown above).

\subsection{Amplitudes extracted from the 6-pion amplitude}
\label{ampl_6pt}

The full six-pion amplitude can be used to extract three-, four- and five-point amplitudes with excited external states via factorization \cite{BrowerChuPRD73}. The amplitudes that we have computed are listed in appendix \ref{ampl_appendix}.

Four-point amplitudes with $\r$ and $\s$ as external states allow to study intermediate states lying on the pion trajectory and its daughters, which were inaccessible from the LS amplitude. Using these four-point amplitudes we find four states at the mass level $\ap m^2 =1$. We compute the three-point amplitudes involving one of them, one pion and one $\s$ or $\r$. These three-point amplitudes are listed in appendix \ref{ampl_appendix}. For the $SU(2)$ flavour symmetry, these four states can be identified with a massive pion $\tilde{\p}$ and three vector mesons $h_1$, $a_1$ and $\w$. In the NS model only the $\w$ meson was present at this mass level ($\ap m^2_{\w,\textup{NS}}=1/2$) that is shifted from the model we here consider by~$1/2$.
These four states should correspond to the mesons $\w(782)$, $h_1(1170)$, $a_1(1260)$ and $\p(1300)$. Taking $m_\r\sim 770 \text{ MeV} = 1/\sqrt{2 \ap}$, the mass of these states is equal to $1/\sqrt{\ap} \sim 1088 \text{ MeV} $ that is an averaged value of their masses.

We also recover the original Veneziano amplitude \cite{Veneziano:1968yb} for the process $\w \to \p\p\p$, which was also present in the NS model. This amplitude can be obtained from the factorization of the six-pion amplitude in the three-particles channel at $\ap s_{123}=1$. One gets the flavour-ordered amplitude
\begin{eqnarray}
A[1_\p,2_\p,3_\p,4_\w]=
2 i \sqrt{C_\textup{6-pt}}
\epsilon^{\mu \nu \rho \sigma} \epsilon_\mu k_{1 \nu }k_{2 \rho} k_{3 \sigma}
\frac{\Gamma [1{\,-\,}\a (s)] \Gamma[1{\,-\,}\a(t)]}{\Gamma[2-\a(s)-\a(t)]} 
\label{O7}
\end{eqnarray}
We have also obtained similar amplitudes involving the mesons $\tilde{\p}$, $a_1$ and $h_1$. They are displayed in Appendix \ref{ampl_appendix}.

Amplitudes extracted from the full six-pion amplitude can help to test its unitarity. In particular, we focus on the pion-sigma scattering amplitude. Assuming $SU(2)$ flavour symmetry, the amplitude can be obtained from the six-pion amplitude factorizing two pairs of pions. In this way one obtains the following result
\begin{equation}
A_{\p\p \s\s}=
-\frac{C_\textup{4-pt}}{4} \delta^{i_1 i_2}
\left\{
\left[(1{+}2 \ap t) B( \tfrac{1}{2}{\,-\,}\ap s,{\,-\,}\ap t)+
(t {\,\leftrightarrow\,}u)\right]+
B({\,-\,}\ap t,{\,-\,}\ap u)
\right\}
\label{eq:A2pi2sigma_short}
\end{equation}
where $B$ is Euler's Beta function. It is amusing to see that the amplitude inherits the Adler zero from the six-pion amplitude. 

The $s$-channel of the previous amplitude is associated to the process $\p \p \to \s \s$, which is not useful to test unitarity since the initial and final states are different.

However, it can be used to clarify the absence of scalar components in the LS amplitude for the scattering $\pi\pi\rightarrow \pi\pi$ at the mass level $\ap s =\frac{3}{2}$ (as shown in Eq. \eqref{eq:LS_res1} choosing $D{\,=\,}4$ and $\a_0{\,=\,} \frac{1}{2}$). Computing the scalar component at $\ap s = \frac{3}{2}$ in Eq. \eqref{eq:A2pi2sigma_short}, we obtain
\begin{equation}
{-}\underset{\ap s=\frac{3}{2},J=0}{\Res} \, A_{2\p 2 \s}
{=} -\frac{C_\textup{4-pt}}{8\ap} G_0(x,4) \d^{i_1 i_2}
\end{equation}
The presence of the non-zero term in the amplitude for the scattering $\pi\pi\rightarrow \sigma\sigma$ implies the existence of (at least) two scalars $\ap s = \frac{3}{2}$ and that one of them is a ghost so that they can exactly cancel each other (opposite residue) in the LS amplitude for the scattering $\pi\pi\rightarrow \pi\pi$  while another combination survives in the $\pi\pi\rightarrow \sigma\sigma$ amplitude \cite{SchwarzPRD72}.  

We turn our attention to the $t$-channel, which is associated to the process $\p \s \to \p \s$. 
The expansion of the residues at the poles $\ap t =n$ (in four spacetime dimensions) of Eq. \eqref{eq:A2pi2sigma_short} gives positive coefficients with the exception of one coefficient at the mass level $\ap t =3$, as shown below. In the following equations we show the expansion of the first residues up to the level $\ap t =3$:
\begin{align}
{-}\underset{\ap t=0}{\Res} \,A_{2\p 2 \s}
{=} & -\frac{C_\textup{4-pt}}{2\ap} G_0(x,4) \d^{i_1 i_2}
\\
{-}\underset{\ap t=1}{\Res} \,A_{2\p 2 \s}
{=} & \,
-\frac{C_\textup{4-pt}}{16\ap}\left[G_1(x,4){+}9 G_0(x,4)\right] \d^{i_1 i_2}
\\
{-}\underset{\ap t=2}{\Res} \,A_{2\p 2 \s}
{=} & \,
-\frac{3C_\textup{4-pt}}{512\ap}\left[27 G_2(x,4){+} 27 G_1(x,4){+} G_0(x,4)\right] \d^{i_1 i_2}
\\
{-}\underset{\ap t=3}{\Res} \,A_{2\p 2 \s}
{=} & \,
-\frac{5C_\textup{4-pt}}{2^{10} 3^3\ap}\left[625 G_3(x,4){+}2125 G_2(x,4)-225 G_1(x,4){+}499 G_0(x,4)\right]\d^{i_1 i_2}
\label{ghost_PiPiSigmaSigma}
\end{align}
Positivity of the coefficients beyond the mass level $\ap s =3$ has been checked up to the mass level $\ap t = 20$ using an algebraic manipulator. 

The presence of states with negative norm in the amplitude $\p \s \to \p \s$ shows that, despite the unitarity of the six-pion flavour-ordered amplitude, the full six-pion amplitude does not satisfy tree-level unitarity.

\section{Adler zeroes and factorisation of the \texorpdfstring{$N$}{N}-pion amplitude}
\label{AdlerNpoint}

In the first part of this section we show that the Adler zeroes that we found in the four and six-pion amplitudes are actually present in all $N$-pion amplitudes, while, in the second part of this section we show that the $N$-pion amplitude factorizes into pion amplitudes, confirming earlier studies \cite{NeveuThornPRL71, SchwarzPRD72} within our approach.

The presence of Adler zeroes can be shown starting from Eqs. \eqref{I12} and \eqref{I13} and taking for instance the limit $k_2 \to 0$. In order to study this limit, it is convenient to parametrize the amplitude with the $N{\,-\,}3$ variables $\a_i$ defined as $z_i= \a_1 \dots \a_{i-2}$ with $i{\,=\,}3, \dots N {\,-\,}1$. It is easy to see  that they are all integrated between $0$ and $1$.

We focus on the terms involving $\a_1$. In the limit $k_2 \to 0$ the terms 
 $(z_2-z_i)^{2\alpha' k_2 k_i}$ with $i>3$ that come from  the Koba-Nielsen term all disappear. The  terms that can give a dependence on $\alpha_1$ are only two kinds of terms, namely $z_2{\,-\,}z_3=1{\,-\,} \a_1$ and $z_i{\,-\,}z_j\propto \a_1$ with $j>i\geq 3$. Therefore in the limit $k_2 \to 0$ the integration over $\a_1$ is decoupled from the integration over the other variables. Let us now see  
 how all this works in more detail.

Let us consider the matrix $A$ shown in Eq. \eqref{I14}. In the limit $k_2 \to 0$ the first row and first column vanish except for the entries $(1,2)$ and $(2,1)$, in which we keep $k_2{\cdot} k_3$ non vanishing because it  will be used as a regulator later on. Therefore, the Pfaffian is reduced to
\begin{equation}
\sqrt{\det A} = \frac{2 \ap k_2{\cdot} k_3 - \frac{1}{2}}{1-\a_1} \sqrt{\det \tilde{A}} \propto
- \frac{\ap s_{23} + \frac{1}{2}}{1-\a_1}\, \a_1^{2-\frac{N}{2}} \sqrt{\det \hat{A}}
\label{pfaffian}
\end{equation}
where the matrix $\tilde{A}$ is obtained from the matrix $A$ by removing the first two rows and columns. In the last step of the previous equation we have used the fact that each non-vanishing element of $\tilde A$ is proportional to $(z_i-z_j)^{-1} \propto \a_1^{-1}$ (with $j>i\geq 3$) and we have introduced the matrix $\hat{A}= \a_1 \tilde{A}$. Passing from the matrix ${\tilde{A}}$ to the matrix ${\hat{A}}$ that does not depend anymore on $\alpha_1$, we get the extra factor $\alpha_1^{2- \frac{N}{2}}$.

From the Koba-Nielsen factor and the additional cyclic factor we can extract the dependence on $\alpha_1$ in the limit of $k_2 \rightarrow 0$, as follows:
\begin{equation}
\lim_{k_2 \rightarrow 0} \prod_{i=2}^N \prod_{i<j} (z_i -z_j)^{2\alpha' k_i k_j} \prod_{i=2}^{N-1} (z_i -z_{i+1})^{- \frac{1}{2}} =
(1-\a_1)^{-\ap s_{23}-\frac{1}{2}} \a_1^{- \ap s_{12}-\frac{N-3}{2}} (\dots)
\end{equation}
where $\left( \dots \right)$ are terms independent on $\a_1$. Collecting all the previous contributions and taking into account that the change of variables from the $z_i$ to the $\alpha_i$ yields a Jacobian factor proportional to $\a_1^{N-4}$, we obtain
\begin{align}
A_N^{\rm NT-S} &= -(\ap s_{23}{+}\frac{1}{2})\int_0^1 d\a_1 \, \a_1^{- \ap s_{12}-\frac{1}{2}} (1-\a_1)^{-\ap s_{23}- \frac{3}{2}} \left[\int_0^1 d \a_2\, \dots \int_0^1 d \a_{N-3}\, (\dots) \right] \nonumber \\
& = \frac{\Gamma(\frac{1}{2}-\ap s_{12})\Gamma(\frac{1}{2}-\ap s_{23})}{\Gamma(-\ap s_{12}-\ap s_{23})}  \left[\int_0^1 d \a_2\, \dots \int_0^1 d \a_{N-3}\, (\dots)  \right]
\end{align}
where $\left(\dots \right)$ are again terms independent on $\a_1$. In the limit of $k_2 \rightarrow 0$ the $\Gamma$-function in the denominator is divergent and therefore, in this limit, the total amplitude vanishes. Since  the $N$-pion amplitude is invariant under cyclic permutations, one can prove the presence of Adler zeroes any time that one of the pion momenta is vanishing.

The factorization of the $N$-pion amplitude into pion amplitudes can be shown as follows. We start again from Eqs. \eqref{I12} and \eqref{I13} and, in order factorise the amplitude at the pole  $s_{I,\dots,N}=-(k_{I}{+}\dots{+}k_N)^2= 0$, where $4\leq I\leq N-2$ is an even number, we consider the limit $z_I \to 0$. The amplitude will be factorised in an $I$-pion and an $(N{\,-\,}I{\,+\,}2)$-pion amplitudes. It is convenient to introduce the variables $x_i$ defined as $z_i=z_I\, x_{i-I+2}$ with $i\geq I$, where $x_2=1$, $x_{N-I+2}=0$. For the remaining $N{\,-\,}I{\,-\,}1$ variables the integration range is $0\leq x_{i+1} \leq x_i$. The change of variables yields the Jacobian $z_I^{N-I-1}$.

The Koba-Nielsen factor and the additional cyclic factor can be split into three factors:
\begin{equation}
\left[\prod_{i=2}^{I-2}\prod_{j=3}^{I-1} (z_i-z_j)^{2 \ap K_i K_j}\right]
\left[\prod_{i=2}^{I-1}\prod_{j=I}^{N} (z_i-z_j)^{2 \ap K_i K_j}\right]
\left[\prod_{i=I}^{N-1}\prod_{j=I+1}^{N} (z_i-z_j)^{2 \ap K_i K_j}\right]
\label{KN_npion}
\end{equation}
$z_I$ appears in the second and third factors. In the limit $z_I \to 0$, the leading term of the second factor yields
\begin{equation}
\prod_{i=2}^{I-1}\prod_{j=I}^{N} (z_i-z_j)^{2 \ap K_i K_j}=
\prod_{i=2}^{I-1} z_i^{2 \ap \sum_{j=I}^N K_i K_j}+\dots=
z_{I-1}^{2 \ap k_{I-1} p- \frac{1}{2}} \prod_{i=2}^{I-2} z_i^{2 \ap k_i p} +\dots
\end{equation}
where the dots are subleading terms in $z_I$ and $p=k_{I}+\dots+k_N$. Together with the first term in Eq. \eqref{KN_npion}, these terms reconstruct the Koba-Nielsen factor and the additional cyclic factor of the $I$-pion amplitude. The third term in Eq. \eqref{KN_npion} yields the Koba-Nielsen factor and the additional cyclic factor of the $(N{\,-\,}I{\,+\,}2)$-pion amplitude:
\begin{equation}
\prod_{i=I}^{N-1}\prod_{j=I+1}^{N} (z_i-z_j)^{2 \ap K_i K_j}=
z_I^{-\ap s_{I \dots N}-\frac{N-I}{2}}\prod_{i=I}^{N-1}\prod_{j=I+1}^{N} (x_{i-I+2}-x_{j-I+2})^{2 \ap K_i K_j}+\dots
\end{equation}
where the dots are again subleading terms in $z_I$. The Pfaffian can be expanded in a similar way and one gets
\begin{equation}
\sqrt{\det A}=\sqrt{\det B} \sqrt{\det C} z_I^{-\frac{N-I}{2}}+\dots
\label{Pfaff_fact}
\end{equation}
where the dots are subleading terms in $z_I$, while the matrices $B$ and $C$ read
\begin{align}
B &= \left( \begin{array}{ccccc}  0  & \frac{2\alpha' K_2 K_3}{z_2-z_3}  & \dots &  \frac{2\alpha' K_2 K_{I-1}}{z_2-z_{I-1}} \\
- \frac{2\alpha' K_2 K_3}{z_2-z_3}  & 0 &  \dots & \frac{2\alpha' K_3 K_{I-1}}{z_3-z_{I-1}} \\
\dots & \dots & \dots & \dots \\
-     \frac{2\alpha' K_2 K_{I-1}}{z_2-z_{I-1}} & -\frac{2\alpha' K_3 K_{I-1}}{z_2-z_{I-1}} & 
 \dots & 0 \end{array} \right)   \\
C &= \left( \begin{array}{ccccc}  0  & \frac{2\alpha' K_I K_{I+1}}{1-x_3}  & \dots &  \frac{2\alpha' K_I K_{N-1}}{1-x_{N-1}} \\
- \frac{2\alpha' K_I K_{I+1}}{1-x_3}  & 0 &  \dots & \frac{2\alpha' K_{I+1} K_{N-1}}{x_3-x_{N-1}} \\
\dots & \dots & \dots & \dots \\
- \frac{2\alpha' K_I K_{N-1}}{1-x_{N-1}} & -\frac{2\alpha' K_{I+1} K_{N-1}}{x_3-x_{N-1}} & 
 \dots & 0 \end{array} \right) 
\end{align}
Notice that, in order to obtain Eq. \eqref{Pfaff_fact}, it is crucial that $I$ is even. If $I$ is odd the matrices $B$ and $C$ have odd dimensions and their determinants vanish.

Collecting all the terms above and integrating over $z_I$ we obtain
\begin{equation}
A^{\rm NT-S}_N(k_1,\dots,k_N) = 
-\frac{C_\textup{N-pt}}{\ap C_{I\textup{-pt}} C_{(N-I+2)\textup{-pt}}}
 A^{\rm NT-S}_I(k_1,\dots,p) \frac{1}{s_{I,\dots N}} A^{\rm NT-S}_{N-I+2}({-}p,\dots,k_N)
\end{equation}
Using this equation we can also fix the normalization $C_\textup{N-pt}$ to be
\begin{equation}
C_\textup{N-pt}= (\ap)^{\frac{N}{2}-2} C_\textup{4-pt}^{\frac{N}{2}-1}
= (\ap)^{-1} \left(-2\p F_\p^2\right)^{1-\frac{N}{2}}
\label{CNpt}
\end{equation}
The factorization of the $N$-pion amplitude into pion amplitudes and the Adler zero should ensure that the field theory limit of the $N$-pion amplitude yields the correspondent amplitude of the non-linear $\s$ model.

The argument goes as follows. Since the normalisation factor $C_\textup{N-pt}$ is proportional to $1/\ap$ while the `world-sheet' integrals are dimensionless, the $\ap$ expansion of the integrals can yield only two kinds of terms, namely terms with a linear dependence on the kinematic invariants ($\ap s_{ij}$) and terms involving poles. The coefficients of the latter terms can be fixed using the factorisation of the $N$-pion amplitude and the field theory limit of pion amplitudes with less than $N$ external legs. The coefficients of the linear terms can be fixed using the Adler zeroes. These coefficients have to cancel the contributions coming from the terms with poles, whose coefficients are found in the previous step.

\section{Eight-pion and four-\texorpdfstring{$\s$}{sigma} amplitudes}
\label{sect:unitarity}

Another check of unitarity can be performed analysing the four-$\s$ amplitude, which can be computed factorizing the eight-pion amplitude given by Eqs. \eqref{I12} and \eqref{I13}. 
In this section we will consider the full $U(N_f)$ flavour group with respect to which $\sigma$ is not a singlet. 
We factorize all the adjacent pion pairs obtaining the following flavour-ordered amplitude
\begin{equation}
A_{4\s}[1,2,3,4]=-\frac{C_\textup{4-pt}}{4}(1{+}\ap s{+} \ap t) \frac{\Gamma(1/2-\ap s)\Gamma(1/2-\ap t)}{\Gamma(1-\ap s-\ap t)}
\end{equation}
Looking at the residues of the full amplitude, we find that at the level $\ap s=5/2$ the coefficients of the expansion in terms of Gegenbauer polynomials can be in general negative:
\begin{align}
{-}\underset{\ap s = 5/2 }{\Res} A_{4\s}=& -\frac{C_\textup{4-pt}}{1920 \ap} 
\Big[-3 (f^{abc})^2 G_3(x,4)+95 (\widehat{d}^{abc})^2 G_2(x,4) \nonumber\\
&-627(f^{abc})^2 G_1(x,4)+535 (\widehat{d}^{abc})^2 G_0(x,4)\Big]
  \qquad (a,b\text{ no sum})
\end{align}
In particular, the negative coefficients are proportional to $(f^{abc})^2$. Furthermore, looking at the next levels, the coefficients of $(f^{abc})^2$ become permanently negative. Therefore, in order to preserve unitarity, $f^{abc}$ must vanish. This is realized only if $\s$ is a singlet of the flavour group, as it is the case for a $U(2)$ flavour group. In this case we have checked the positivity of the coefficients up to the level $\ap s =19/2$. We show the first three levels:
\begin{align}
{-}\underset{\ap s = 1/2 }{\Res} A_{4\s}=&\,-\frac{9C_\textup{4-pt}}{8 \ap} G_0(x,4)\\
{-}\underset{\ap s = 3/2 }{\Res} A_{4\s}=&\,-\frac{C_\textup{4-pt}}{48 \ap} [G_2(x,4)+25 G_0(x,4)] \\
{-}\underset{\ap s = 5/2 }{\Res} A_{4\s}=& -\frac{C_\textup{4-pt}}{384 \ap} 
\Big[19 G_2(x,4)+107 G_0(x,4)\Big]
\end{align}

In comparison with the amplitude $A_{4 \s}$ (in the case $U(N_f)$), the ghost already found in the $A_{\p\p\s\s}$ amplitude is harder to expose because of the larger positive couplings (residues) of the states with positive norm. In the amplitude $A_{4 \s}$ instead, ghosts start to appear systematically at each mass level. The  difference between these amplitudes may be due to the fact that $A_{4 \s}$ is derived by factorisation of an eight-pion amplitude, which is not `expected' to be unitary to start with even for a given flavour-ordering, while the LS and $A_{\p\p\s\s}$ amplitudes can be constructed using four- and six-pion flavour-ordered amplitude that are unitary as we have shown.  

\section{Conclusions and outlook}
\label{conclusions}

With a slight modification of the $N$-tachyon amplitude of the NS model an $N$-pion amplitude with massless pions can be constructed \cite{NeveuThornPRL71, SchwarzPRD72}. The four-pion amplitude is equal to the one of the Lovelace-Shapiro model \cite{LOVELACE, SHAPIRO} that seems to be free of  ghosts for the space-time dimension $D=4$, has Adler zeroes and reproduces in the field theory limit ($\alpha' \rightarrow 0$ with fixed $F_\pi$)  the four-pion amplitude of the non-linear $\sigma$-model. We have then reconsidered $N$-point pion amplitudes that were known to have Adler zeroes and we have shown that they reproduce the $N$-point pion amplitudes of the non-linear $\sigma$-model in the field theory limit. 

The $N$-pion amplitude we considered is obtained from the $N$-tachyon amplitude of the NS model by shifting the tachyon mass to zero. With such a shift the massless gauge boson gets a non-zero mass becoming the  $\rho$-meson and what is colour symmetry in the NS model becomes a flavour symmetry in the model we consider here. In this way the super-conformal invariance of the NS model is of course lost, but the integrand of the amplitude keeps the super-projective invariance. This implies that, in general, the model has negative norm states  (ghosts).  On the other hand, it  gives a reasonable overall description   of   the low-energy $\pi \pi$ scattering although the mass of the various excited states are not always quite agreeing with the experimental data. 

We have also shown that flavour-ordered four and six-pion amplitudes  can be derived from the correspondent tachyon amplitudes in the ten-dimensional NS model by suitably choosing the KK momenta of the tachyons along the six extra/compactified directions, using a different but equivalent choice from the early studies \cite{NeveuThornPRL71, FairlieNPB72, SchwarzPRD72, SinghPasupathyPRD74, BrowerChuPRD73}. As a result we have shown  that these flavour-ordered amplitudes satisfy tree-level unitarity, an issue largely overlooked in \cite{NeveuThornPRL71, FairlieNPB72, SchwarzPRD72, SinghPasupathyPRD74, BrowerChuPRD73}. We have then carefully studied their properties and derived scattering amplitudes involving 
$\rho$ and $\omega$ mesons as well as $\sigma$ by factorisation of higher point pion amplitudes.

The model we considered has exactly linearly rising Regge trajectories. As discussed in Ref.~\cite{CKSZ}, this is fine in the unphysical region where both $s$ and $t$ are positive, but, according to asymptotic freedom in QCD, one expects that the Regge trajectory  will level down for negative values of the Mandelstam variables.  In order to solve this problem and also to have a power-like  behaviour at large $s$ and $|t|$, as also predicted by QCD,  it has been proposed in Refs.~\cite{AS,VYO}  to sum over an infinite number of  four-point amplitudes   with Regge slopes ${\alpha'}/{n}$.  Although more complicated, this procedure can also be applied to the higher-point amplitudes and, as discussed in Ref.~\cite{VYO}, could   provide a connection with  holographic 
calculations~\cite{Witten:1998zw,Sakai:2004cn,Bartolini:2016dbk,BPST,AI}.
 Another holography-inspired approach to hadron scattering has been proposed that rely on open strings with massive ends \cite{Sonnenschein:2016pim, Sonnenschein:2017ylo, Sonnenschein:2018fph, Sonnenschein:2019bca}. It would be very interesting to compare our results for pion scattering and those that would emerge from \cite{Sonnenschein:2019bca} in some chiral limit. Another complementary approach relies on the S-matrix bootstrap for QCD that has been recently revived in the context of $\pi\pi$ scattering in \cite{Guerrieri:2019rwp}. The  parameter space of chiral zeros, scattering lengths, and resonance masses has been carefully explored in \cite{Guerrieri:2019rwp} and the remarkable location where QCD seems to lie has been tightly constrained. It would be interesting to compare these results with those obtained in our approach.


\acknowledgments

We acknowledge fruitful discussions with Alice Aldi, Andrea Addazi, Richard Brower, Ram Brunstein, Davide Bufalini, Maurizio Firrotta, Roberto Frezzotti, Francesco Fucito, Michael Green, Alfredo Grillo, Andrea Guerrieri, Zohar Komargodski, Raffaele Marotta, Francisco Morales, Giancarlo Rossi, Rodolfo Russo, Alberto Salvio, Cobi Sonnenschein, Nazario Tantalo, Gabriele Veneziano, Congkao Wen and the late Yassen Stanev. We especially thank Oliver Schlotterer, Cobi Sonnenschein and Gabriele Veneziano for discussions and for a critical reading of the manuscript.
DC, MB and PDV  would like to thank the Galileo Galilei Institute for hospitality during the workshop
``String theory from the worldsheet perspective'' where they had the opportunity to carry part of this work.  
A preliminary version of this work was presented at the {\it ``String Geometry and String Phenomenology Institute"} held at CERN.  MB would like to thank the organisers  for their kind hospitality and the participants for their comments and queries.
This work is partially supported by the grant {\it ``Strong Interactions: from Lattice QCD to Strings, Branes and Holography''} (CUN Area 02, CUP E84I19002260005), within the scheme {\it ``Beyond Borders''} of the University of Roma ``Tor Vergata".
DC was supported by FWF Austrian Science Fund via the SAP P30531-N27.
PDV was supported as a Simons GGI scientist
from the Simons Foundation grant 4036 341344 AL. The research of PDV
is partially supported by the Knut and Alice Wallenberg Foundation
under grant KAW 2018.0116.


\appendix

\section{Super-projective transformations}
\label{A}
A super-projective transformation acts on the variable $Z\equiv (z,\theta)$ as follows:
\begin{eqnarray}
z \rightarrow z' \equiv \frac{Az+B +\alpha \theta}{Cz+D + \beta \theta}~~;~~ \theta \rightarrow  \theta' \equiv \frac{{\bar{\alpha}} z + {\bar{\beta}} + {\bar{A}} \theta}{Cz+D + \beta \theta}~`;~~ AD-BC=1
\label{NSM1}
\end{eqnarray}
where
\begin{eqnarray}
{\bar{A}} = 1 -\frac{3}{2} \alpha \beta ~~;~~{\bar{\alpha}}=- C\alpha + A \beta~~;~~ {\bar{\beta}} = - D \alpha + B \beta
\label{NSM2}
\end{eqnarray}
From it, it is easy to derive the following transformations:
\begin{eqnarray}
Z_i -Z_j \Longrightarrow \frac{(Z_i -Z_j) (1- \alpha \beta)}{(Cz_i +D + \beta \theta_i) Cz_j +D + \beta \theta_j) }
\label{NSM3}
\end{eqnarray}
and
\begin{eqnarray}
dZ_i \Longrightarrow dZ_i \frac{1- \frac{1}{2} \alpha \beta}{Cz_i +D + \beta \theta_i}  
\label{NSM4}
\end{eqnarray}
This last transformation is obtained by computing the super-determinant of the matrix
\begin{eqnarray}
\left( \begin{array}{cc} \frac{\partial z'}{\partial z} &  \frac{\partial z'}{\partial \theta}\\
\frac{\partial \theta'}{\partial z} & \frac{\partial \theta'}{\partial \theta} \end{array} \right)
=  \left( \begin{array}{cc} \frac{1 + (A\beta -C\alpha) \theta}{(Cz +D + \beta \theta)^2} & \frac{\alpha (Cz+D)- \beta (Az+B) + 2\alpha \beta \theta}{(Cz +D + \beta \theta)^2} \\
\frac{\beta - C(1- \frac{1}{2} \alpha \beta)\theta}{(Cz +D + \beta \theta)^2} & 
\frac{(Cz+D) \left(1- \frac{1}{2} \alpha \beta \right) +2 \beta \theta}{(Cz +D + \beta \theta)^2} \end{array} \right)
\equiv \left( \begin{array}{cc} A_{11} & A_{12} \\ A_{21} & A_{22} \end{array} \right)
\label{NSM5}
\end{eqnarray}
that is equal to
\begin{eqnarray}
\frac{A_{11} - A_{12} A_{22}^{-1} A_{21}}{A_{22}} = 
\frac{1 - \frac{1}{2} \alpha \beta}{Cz +D + \beta \theta} 
\label{NSM6}
\end{eqnarray}

\section{Explicit computation of the \texorpdfstring{$N$}{N}-point amplitude}
\label{explicit}

In this Appendix we  perform the integration over the variables $\theta_i$  for  the $N$-point amplitude in Eq. (\ref{I11}).  Taking  advantage of the superprojective invariance we fix $z_1 =\infty$, $z_2=1$, $z_N =0$ and  $\theta_1 = \theta_N=0$. In this case the integrand in Eq. (\ref{I11}) can be written as follows:
\begin{eqnarray}
&&  \prod_{i=2}^N \prod_{i<j} (z_i -z_j)^{2\alpha' k_i k_j} \prod_{i=2}^{N-1} (z_i -z_{i+1})^{- \frac{1}{2}} \Bigg( 1 - \theta_2 \sum_{j=3}^{N-1} \theta_j \frac{2 \alpha' K_2 K_j}{z_2 - z_j} \Bigg)\times \\
&&  \Bigg( 1 - \theta_3 \sum_{j=4}^{N-1} \theta_j\frac{2 \alpha' K_3 K_j}{z_3 - z_j} \Bigg)
\Bigg( 1 - \theta_4 \sum_{j=5}^{N-1} \theta_j\frac{2 \alpha' K_4 K_j}{z_4 - z_j} \Bigg) \dots 
\Bigg( 1 - \theta_{N-2}  \frac{2 \alpha' K_{N-2} K_{N-1} }{ z_{N-2} - z_{N-1} } \Bigg) 
\nonumber 
\label{EC1}
\end{eqnarray}
where 
\begin{eqnarray}
&& K_i K_j = k_i k_j -\frac{1}{4\alpha'} ~~~~\text{ if}~~~~j=i\pm 1 ~~;~~K_i K_j = k_i k_j ~~\text{if}~~~j \neq i\pm 1
\label{EC2}
\end{eqnarray}
The integration over the variables $\theta_i$ of the previous expression is performed by collecting all terms that have the product of all $\theta_i$. The simplest way to collect all these terms is by taking 
the term with $\theta_2 \theta_3$ in the first round bracket, the term with $1$ in the second round bracket, the term with $\theta_4 \theta_3$ in the third round bracket and so on. In this case one gets:
\begin{eqnarray}
&&  \prod_{i=2}^N \prod_{i<j} (z_i -z_j)^{2\alpha' k_i k_j} \prod_{i=2}^{N-1} (z_i -z_{i+1})^{- \frac{1}{2}} 
(-1)^{\frac{N}{2}+1}  \theta_2 \dots \theta_{N-1}
\prod_{i=1}^{(N-2)/2} \frac{ 2\alpha' K_{2i} K_{2i+1}}{z_{2i} - z_{2i+1}}
\label{EC3}
\end{eqnarray}
Then we have to compute the following integral:
\begin{eqnarray}
\int d\theta_2 \int d\theta_3 \dots \int d\theta_{N-1} \theta_2 \theta_3 \dots  \theta_{N-1} = 
(-1)^{1+ \frac{N}{2}(N-5)} \prod_{i=2}^{N-1} \int d\theta_i \theta_i  = (-1)^{\frac{N+2}{2}}
\label{EC4}
\end{eqnarray}
where we have used the relation $\int d \theta_i \theta_i=1$.

In conclusion, the amplitude corresponding to the first of the $(N-3)!!$ terms is given by:
\begin{eqnarray}
\prod_{i=3}^{N-1} \int d z_i  \prod_{i=2}^N \prod_{i<j} (z_i -z_j)^{2\alpha' k_i k_j} \prod_{i=2}^{N-1} (z_i -z_{i+1})^{- \frac{1}{2}} \prod_{i=1}^{(N-2)/2} \frac{ 2\alpha' K_{2i} K_{2i+1}}{z_{2i} - z_{2i+1}}
\label{EC5}
\end{eqnarray}
One can compute in an analogous way the remaining terms. It turns out that the final result can be written in the following compact way:
\begin{eqnarray}
A_N^{\rm new } = \prod_{i=3}^{N-1} \int d z_i  \prod_{i=2}^N \prod_{i<j} (z_i -z_j)^{2\alpha' k_i k_j} \prod_{i=2}^{N-1} (z_i -z_{i+1})^{- \frac{1}{2}}  \sqrt{\det A}
\label{EC6}
\end{eqnarray}
where $A$ is the  antisymmetric matrix in Eq. (\ref{I13}).

In the case of $N=4$  we get
\begin{eqnarray}
\sqrt{\det A} = \frac{2\alpha' K_2 K_3}{z_2-z_3}
\label{EC7}
\end{eqnarray}
while in the case of $N=6$ we get
\begin{eqnarray}
\sqrt{\det A} = \frac{2\alpha' K_2 K_3}{z_2-z_3}\frac{2\alpha' K_4 K_5}{z_4-z_5}+ \frac{2\alpha' K_2 K_5}{z_2-z_5}
\frac{2\alpha' K_3 K_4}{z_3-z_4} - \frac{2\alpha' K_2 K_4}{z_2-z_4}\frac{2\alpha' K_3 K_5}{z_3-z_5}
\label{EC8}
\end{eqnarray}

\section{Group-theoretical properties of the generators of \texorpdfstring{$U(N_f)$}{U(Nf)}}
\label{group}

The generators $T^a$ of $U(N_f)$ that enter in the Chan-Paton factors satisfy the following properties:
\begin{eqnarray}
[ T^a , T^b] = i f^{abc} \sqrt{2} T^c~;~\{T^a , T^b \} = \sqrt{2} {\hat{d}}^{abc}~;~{\rm Tr} (T^a T^b) = \delta^{ab}~~;~~a,b,c = 1 \dots N_f
\label{U3}
\end{eqnarray}
where $f^{abc}$ are the structure constants of $U(N_f)$ and ${\hat{d}}^{abc}$ are the symmetric invariant tensors of $U(N_f)$. They are obtained through the following relations:
\begin{eqnarray}
{\rm Tr} ( T^a [T^b , T^c]) = i \sqrt{2} f^{abc}\quad , \quad  {\rm Tr} ( T^a \{T^b , T^c \}) =  \sqrt{2} {\hat{d}}^{abc}
\label{U4}
\end{eqnarray}
In $U(N_f)$ we have also the $U(1)$ generator $T^0 = {\idnty}/{\sqrt{N_f}}$ and ${\hat{d}}^{abc} = d^{abc}$ if all the indices represent $SU(N_f)$ generators, while ${\hat{d}}^{0ab} = \sqrt{{2}/{N_f}} \delta^{ab}$  if one of the indices correspond to the $U(1)$ generator.

For $A$ and $B$ arbitrary matrices, the Chan-Paton factors satisfy the following relation:
\begin{eqnarray}
\sum_a {\rm Tr} ( A T^a) {\rm Tr}(T^a B) = {\rm Tr}(AB)
\label{CPrelation}
\end{eqnarray}
that is useful to study the factorization properties of the residues at the poles.

\section{Linear and non-linear sigma model}
\label{linear}

In this Appendix we review the relevant scattering amplitudes in the linear and non-linear $\s$-models to be compared with the amplitudes computed in this paper from our proposed model.

The Lagrangian of the $SO(4)$ linear $\sigma$-model is equal to
\begin{equation}
\calL= -\frac{1}{2} (\der \vec{\P})^2-\frac{1}{2} (\der s)^2-V(\vec{\P},s)
\label{lagra}
\end{equation}
where $\vec{\P} = (\Pi^1, \Pi^2,\Pi^3)$ and the potential reads
\begin{equation}
V(\vec{\P},s)=\frac{\l}{4} (\vec{\P}^2+s^2)^2+\frac{\m^2}{2} (\vec{\P}^2+s^2)  + c\, s
\label{eq:unbroken_potential}
\end{equation}
The first two terms of the potential are $SO(4)$ invariant, while the last term breaks explicitly the $SO(4)$ symmetry. If $\mu^2 <0$ the minimum of the potential requires a non-vanishing vacuum expectation value for the field $s$: $\<s\>=v$ that spontaneously breaks the $SO(4)$ symmetry. The minimum of the potential is given by $c+v(\m^2+ \l v^2)=0$. After the breaking of the $SO(4)$ symmetry, the vev $v$, the masses of the fluctuation $\s=s-v$ and $\P$ can be used instead of $\l$, $\m^2$ and $c$ to parametrise the potential. We get
\begin{equation}
V(\P,\s)= \frac{m_\s^2}{2} \s^2+\frac{m_\p^2}{2} \vec{\P}^2+\frac{\D m^2}{2v}\s \vec{\P}^2+\frac{\D m^2}{2v}\s^3+\frac{\D m^2}{8v^2} \s^4+\frac{\D m^2}{4v^2} \s^2 \vec{\P}^2+\frac{\D m^2}{8v^2} (\vec{\P}^2)^2
\label{eq:broken_potential}
\end{equation}
where $\D m^2{\,=\,}m_\s^2{\,-\,}m_{\P}^2$.  From the Lagrangian in \eqref{lagra} with the previous potential we can compute the two three-point amplitudes
\begin{equation}
A_{\p \p \s}^{\text{(L}\sigma \text{M)}}=- \frac{\D m^2}{v} \d^{i_1 i_2}
\quad , \quad
A_{\s \s \s}^{\text{(L}\sigma \text{M)}}=- \frac{3\D m^2}{v}
\end{equation}
and the three four-point amplitudes
\begin{align}
A_{\s\s\s\s}^{\text{(L}\sigma \text{M)}}&=-\frac{3 \D m^2}{v^2} \left[1+ \frac{3 \D m^2}{s-m_\s^2}+ \frac{3 \D m^2}{t-m_\s^2}+ \frac{3 \D m^2}{u-m_\s^2}  \right]\\
A_{\s\s\p\p}^{\text{(L}\sigma \text{M)}}&=-\frac{\D m^2}{v^2}\d^{i_3 i_4}\left[3 \,\frac{s-m_\p^2}{s-m_\s^2}+ \frac{t-m_\s^2}{t-m_\p^2}+ \frac{u-m_\s^2}{u-m_\p^2}\right]\\
A_{\p\p\p\p}^{\text{(L}\sigma \text{M)}}&=-\frac{\D m^2}{v^2}\left[ \frac{s-m_\p^2}{s-m_\s^2}\d^{i_1 i_2}\d^{i_3 i_4}+\frac{t-m_\p^2}{t-m_\s^2}\d^{i_1 i_4}\d^{i_2 i_3} +\frac{u-m_\p^2}{u-m_\s^2}\d^{i_1 i_3}\d^{i_2 i_4} \right]
\label{eq:APiPiPiPi_LSM}
\end{align}
The last two amplitudes have Adler zeroes. The flavour-ordered amplitude associated to the last amplitude reads
\begin{equation}
A^{\text{(L}\sigma \text{M)}}[1,2,3,4]=-\frac{\D m^2}{2v^2}\left[ \frac{s-m_\p^2}{s-m_\s^2}+\frac{t-m_\p^2}{t-m_\s^2}\right]
\label{eq:Apppp_CO_LSM}
\end{equation}

Let us now consider a $SU(N)$ non-linear $\s$-model given by the lagrangian
\begin{align}
\calL=&\frac{F_\p^2}{4} \Tr[U^{-1} \der_\m U U^{-1} \der^\m U] +\frac{m_\p^2 F_\p}{2 \sqrt{2}} \Tr [U+U^\dagger]   \\
&U(\P_a)= \frac{F_\p}{\sqrt{2}} \exp \left( i\frac{ \sqrt{2}}{F_\p}\, \P^a\, T_a \right)
\end{align}
Expanding the exponentials, all the terms with an odd number of pions cancel. The first interaction term of the expansion yields a quartic interaction. Therefore the first non-trivial flavour-ordered amplitude is the four-point amplitude
\begin{align}
A^{\text{(NL}\sigma \text{M)}}[1,2,3,4]=\frac{s+t-2 m_\p^2}{2 F_\p^2}
\label{eq:NLSM_4pt_color_ordered}
\end{align}
The previous amplitude vanishes at $s{\,=\,}t{\,=\,}m_\p^2$. When we have $SU(2)$ as flavour group the full amplitude can be written as
\begin{equation}
A^{\text{(NL}\sigma \text{M)}}_{4\p}=
\delta^{i_1 i_2}\delta^{i_3 i_4}\frac{s{-}m_\pi^2}{F_\p^2}+
\delta^{i_1 i_3}\delta^{i_2 i_4}\frac{u{-}m_\pi^2}{F_\p^2}+
\delta^{i_1 i_4}\delta^{i_2 i_3}\frac{t{-}m_\pi^2}{F_\p^2}
\label{eq:APiPiPiPi_LSM_Limit}
\end{equation}
This amplitude matches the amplitude \eqref{eq:APiPiPiPi_LSM} in the limit $m_\s^2 \to \infty$, where the $\s$ meson is integrated out, if we choose
\begin{equation}
v^2=F_\p^2
\label{eq:linearVSnonlinear}
\end{equation}
The next non-trivial amplitude is the six-point amplitude given by
\begin{align}
A^{\text{(NL}\sigma \text{M)}}_{6\p}&[1,2,3,4,5,6]=
\frac{1}{4 F_\p^4} \Bigg[s_{12}{+}s_{23}{+}s_{34}{+}s_{45}{+}s_{56}{+}s_{16}+
\nonumber \\
&{-}\frac{(s_{12}{+}s_{23})(s_{34}{+}s_{45})}{s_{123}}
 {-}\frac{(s_{23}{+}s_{34})(s_{45}{+}s_{56})}{s_{234}}
 {-}\frac{(s_{34}{+}s_{45})(s_{56}{+}s_{16})}{s_{345}}
 \Bigg]
\label{eq:6ptNLSM}
\end{align}

\section{Six-pion amplitude field theory limit}
\label{field_theory_limit}

In this Appendix we give some detail on the calculation of
the field theory limit of the six-pion flavour-ordered amplitude \eqref{eq:paolo_6pt}. The general strategy can be summarized as follows: first we write the amplitude in a more convenient parametrization, then we expand the amplitude using the binomial series,  compute the world-sheet integrals and pick up the relevant terms for the leading  term in the $\ap$-expansion and finally we re-sum the series. In the $\ap$-expansion of the world-sheet integrals we keep terms up to $\ap^1$ since $C_\textup{6-pt} \sim \ap^{-1}$.

Let us start introducing a different parametrization\footnote{For more details on this parametrization and its application to tree-level amplitude of $N$ massless open strings see Ref. \cite{Mafra:2011nw}.} for the amplitude in Eq. \eqref{eq:paolo_6pt} which manifestly shows all the planar channels. We introduce six variables $X_i$ associated to the two-particle channels $(i,i{\,+\,}1)$, with $i=1,\dots,6$, and three variables $Y_j$, with $j=1,2,3$, related to the three-particle channels $(j,j{\,+\,}1,j{\,+\,}2)$. When one of these variables vanishes the poles in the correspondent channel are made manifest.

In  Eq. \eqref{eq:paolo_6pt}, the variables $\a$, $\b$ and $\g$ can be directly identified with $X_1$, $Y_1$ and $X_5$ respectively. The terms which are not a power of one of these three variables, like $(1{\,-\,}\dots)$, can be directly replaced by products of the new variables as follow
\begin{align}
(1-\alpha)&= X_2 X_6 Y_2 \qquad 
(1-\beta) = X_3 X_6 Y_2 Y_3 \qquad 
(1-\gamma) = X_4 X_6 Y_3 \\
(1-\alpha \beta) & = X_6 Y_2 \qquad
~~(1-\beta \gamma) = X_6 Y_3 \qquad
~~~(1-\alpha \beta \gamma) = X_6
\end{align}
The logic of these replacements is related to which channel each single term contributes. For instance the term $(1{\,-\,}\alpha)$ can contribute to the channels $(2,3)$ ($\a \to 1$), $(2,3,4)$ ($\a,\b \to 1$) and $(1,6)$ ($\a,\b,\g \to 1$), thus $(1-\alpha) = X_2 Y_2 X_6$.

Introducing six new variables we have to insert six delta-functions. Their presence is necessary to avoid simultaneous poles in incompatible channels. These delta-functions insure that, when one of variables vanishes, the variables associated to incompatible channels goes to one. In this way one  avoids the simultaneous presence of poles in incompatible channels.
Since our starting amplitude in Eq. \eqref{eq:paolo_6pt} is already written in terms of the variables $X_1$, $X_5$ and $Y_1$, the constraints for the remaining variables are the following:
\begin{align}
\Delta = \,&
\delta(X_2+X_1 X_3 Y_3 - 1)
\delta(X_3+X_2 X_4 Y_1 - 1)
\delta(X_4+X_3 X_5 Y_2 - 1)\times \nonumber\\
\times \, & \delta(X_6+X_1 X_5 Y_1 - 1)
\delta(Y_2+X_1 X_4 Y_1 Y_3 - 1)
\delta(Y_3+X_2 X_5 Y_1 Y_2 - 1)
\label{delta_incomp}
\end{align}
To complete the change of variable we have to multiply the integrand by the factor coming from the integration measure
\begin{equation}
d\a\,d\b\,d\g =
(1{\,-\,}X_1 Y_1)(1{\,-\,}X_5 Y_1)\, \D \prod_{i=1}^6 dX_i \prod_{j=1}^3\, dY_j = 
X_6^2 Y_2 Y_3 \,\D \prod_{i=1}^6 dX_i \prod_{j=1}^3\, dY_j 
\end{equation}
Collecting all the ingredients together we obtain
\begin{align}
A_6=\,& C_\textup{6-pt}
\prod_{i=1}^6 \int_0^1  dX_i\, X_i^{-\alpha' s_{i\,i+1} - \frac{1}{2}} \prod_{i=j}^3 \int_0^1dY_j \, Y_j^{-\alpha' s_{i\,i+1\,i+2} } \times \Delta
\nonumber \\
& \times  
\left[
\frac{\left(\alpha' s_{34}+\frac{1}{2} \right)\left( \alpha' s_{56}+\frac{1}{2} \right)}{X_1 X_3 X_5 Y_1 Y_2 Y_3}
-\frac{ \alpha'^2 s_{35} s_{46} }{X_1 Y_1 Y_3} + \frac{\alpha's_{36} \left(\alpha's_{45}+\frac{1}{2} \right)}{X_1 X_4 Y_1 Y_3} 
\right]
\label{eq:paolo_6pt_XY}
\end{align}
In order to go back to the previous representation involving $X_1$, $Y_1$ and $X_5$ one must solve the equations required by the delta-functions in Eq \eqref{delta_incomp} obtaining 
the other variables in terms of  $X_1$, $Y_1$ and $X_5$
\begin{gather}
X_2= \frac{1 - X_1}{1 - X_1 Y_1} \quad , \quad
X_3 =\frac{(1 - Y_1) (1 - X_1 X_5 Y_1)}{(1 - X_1 Y_1) (1 - X_5 Y_1)}  \quad , \quad
X_4 = \frac{1 - X_5}{1 - X_5 Y_1} 
\label{eq:inverse_X1X5Y1} \\
X_4 =\frac{1 - X_5}{1 - X_5 Y_1}\quad, \quad
X_6 = 1 - X_1 X_5 Y_1 \quad , \quad
Y_2 = \frac{1 - X_1 Y_1}{1 - X_1 X_5 Y_1} \quad , \quad
Y_3 = \frac{1 - X_5 Y_1}{1 - X_1 X_5 Y_1} \nonumber
\end{gather}
Notice that the incompatibility relations for the variables $X_1$, $X_5$ and $Y_1$ (for instance $Y_1+X_3 X_6 Y_2 Y_3=1$), which are not directly imposed by the explicit delta functions, are also valid. One can prove them combining the relations in Eq. \eqref{eq:inverse_X1X5Y1}.

Using the six delta-functions we can choose to have ($X_2$,$X_6$,$Y_2$) or ($X_1$,$X_3$,$Y_3$) as independent variables instead of ($X_1$,$X_5$,$Y_1$) as in Eq. \eqref{eq:paolo_6pt}. For the first choice we get
\begin{align}
A_6^{X_2X_6Y_2}&{\,=\,} C_\textup{6-pt} \!
\int_0^1 \!dX_2\, dY_2 \, dX_6 
X_2^{-\alpha' s_{23} - \frac{1}{2}}
(1{-}X_2)^{-\alpha' s_{34} - \frac{1}{2}}
Y_2^{-\alpha' s_{234} }
(1{-}Y_1)^{ - \alpha' s_{45} - \frac{1}{2}} 
\label{eq:paolo_6pt_X2X6Y2} \\
& \times X_6^{ -\alpha' s_{16} - \frac{1}{2}}
(1{-}X_6)^{-\alpha' s_{56} - \frac{1}{2} }
(1{-}X_2 Y_2)^{-\alpha' s_{35}}
(1{-}X_6 Y_2)^{-\alpha' s_{46}} 
(1{-}X_2 X_6 Y_2)^{ -\alpha' s_{36} } 
\nonumber \\
& \times  
\left[
\frac{ \left(\alpha' s_{34}+\frac{1}{2} \right)\left( \alpha' s_{56}+\frac{1}{2} \right) }{(1-X_2)(1-X_6)Y_2}
-\frac{ \alpha'^2 s_{35} s_{46} }{(1-X_2 Y_2)(1-X_6 Y_2)} + \frac{\alpha's_{36} \left(\alpha's_{45}+\frac{1}{2} \right)}{(1-X_2 X_6 Y_2) (1-Y_2)}
\right]
\nonumber
\end{align}
Notice that this amplitude has the same structure of the amplitude in Eq. \eqref{A6igig} up to a cyclic permutation. The second choice of variables, ($X_1$,$X_3$,$Y_3$), yields instead
\begin{align}
A_6^{X_1X_3Y_3}=\,&  C_\textup{6-pt} 
\int_0^1 dX_1\, dY_3 \, dX_3
X_1^{-\alpha' s_{12} - \frac{1}{2}}
(1{-}X_1)^{-\alpha' s_{16} - \frac{1}{2}}
Y_3^{-\alpha' s_{345} }
(1{-}Y_3)^{ - \alpha' s_{56} - \frac{1}{2}} 
\nonumber \\
& \times \! X_3^{ -\alpha' s_{34} - \frac{1}{2}}
(1{-}X_3)^{-\alpha' s_{45} - \frac{1}{2} }
(1{-}X_1 Y_3)^{-\alpha' s_{15}}
(1{-}X_3 Y_3)^{-\alpha' s_{46}} 
(1{-}X_1 X_3 Y_3)^{ -\alpha' s_{14} } 
\nonumber \\
& \times  
\left[
\frac{ \left(\alpha' s_{34}+\frac{1}{2} \right)\left( \alpha' s_{56}+\frac{1}{2} \right) }{X_1 X_3(1-Y_3)Y_3}
-\frac{ \alpha'^2 s_{35} s_{46} }{X_1 Y_3(1-X_3 Y_3)} + \frac{\alpha's_{36} \left(\alpha's_{45}+\frac{1}{2} \right)}{X_1 Y_3 (1-X_3)}
\right]
\label{eq:paolo_6pt_X1X3Y3}
\end{align}
which has a structure similar to Eq. \eqref{eq:paolo_6pt}.

The amplitude in Eq. \eqref{eq:paolo_6pt_XY} shows the presence of massless and (appearently) tachyonic poles in incompatible three- and two-particle channels respectively, which can lead to ambiguous divergences in the $\ap$-expansion. The new parametrizations allow us to manipulate the integrand in order to only have massless poles in compatible channels. This follows from the incompatibility relations in Eq.  \eqref{delta_incomp}. 
For instance, in the second term of Eq. \eqref{eq:paolo_6pt_XY}, there are massless poles in the channels $(1,2,3)$ and $(3,4,5)$, which are not compatible. Using the relation $Y_3+X_2 X_5 Y_1 Y_2=1$, we can separate them as follow 
\begin{equation}
\frac{1}{X_1 Y_1 Y_3}=\frac{1}{X_1 Y_1}+\frac{X_2 X_5 Y_2}{X_1 Y_3}
\end{equation}
Applying this procedure to all the three integrals in Eq. \eqref{eq:paolo_6pt_XY} we get
\begin{align}
\frac{1}{X_1 X_3 X_5 Y_1 Y_2 Y_3} &=
\frac{1}{X_1 X_3 X_5}{+}\frac{1}{X_1 X_5 Y_1}{+}\frac{1}{X_3 X_5 Y_2}{+}\frac{1}{X_1 X_3 Y_3}{+}\nonumber\\
&~~~{-}\frac{1}{X_1}{-}\frac{1}{X_3}{-}\frac{1}{X_5}{+}X_2 X_4 X_6 Y_1 Y_2 Y_3
\label{eq:decompositionI1}
\\
\frac{1}{X_1 Y_1 Y_3}&=\frac{1}{X_1 Y_1}{+}\frac{1}{X_1 Y_3}{-}\frac{X_4}{X_1}{-}X_3 X_5 Y_2 
\label{eq:decompositionI2}\\
\frac{1}{X_1 X_4 Y_1 Y_3}&= 
\frac{1}{X_1 X_4 Y_1}{+}\frac{1}{X_1 X_4 Y_3}{-}\frac{1}{X_1}{-}\frac{1}{X_4}{+}1
\label{eq:decompositionI3}
\end{align}
Some of these integrals are related by cyclic transformations or reflections. Therefore, in order to compute the limit $\ap \rightarrow 0$ it is enough to only consider the following seven 
integrands
\begin{equation}
\frac{1}{X_1 X_3 X_5} ~~, ~~
\frac{1}{X_1 X_4 Y_1} ~~, ~~
\frac{1}{X_1 X_5 Y_1} ~~, ~~
\frac{1}{X_1 Y_1} ~~, ~~
\frac{1}{X_1} ~~, ~~
1 ~~, ~~
X_2 X_4 X_6 Y_1 Y_2 Y_3
\label{eq:relevant_integrals}
\end{equation}
In 
Eq. \eqref{eq:decompositionI2} we keep only $\frac{1}{X_1 Y_1}$  because the numerator is proportional to $\alpha'^2$ and there is a massless pole. Thus this is the only term in Eq.  \eqref{eq:decompositionI2} that can produce a simple power of $\alpha'$.

All the chosen relevant integrals in Eq. \eqref{eq:relevant_integrals} have a massless pole in the channel $(1,2,3)$ or none. Therefore, we can use the variables $(X_1,X_5,Y_1)$ and integrate over all the other variables using the delta-functions. For each integral the $\alpha'$-expansion generates several terms which have to be summed. In the following we will explicitly show only the simplest non-trivial case.

\subsection{An example: \texorpdfstring{$1/ X_1 X_5 Y_1$}{1/X1 X5 Y1}}

The integral that we want to expand, up to order $\ap^1$, is the following
\begin{align}
I_{X_1X_5Y_1}&= \int_0^1 \prod_{i=1}^6  dX_i\, X_i^{-\alpha' s_{i\,i+1} - \frac{1}{2}} \prod_{i=j}^3 dY_j \, Y_j^{-\alpha' s_{i\,i+1\,i+2} } \frac{\Delta}{X_1 X_5 Y_1} = \label{example_integral}\\
& =
\int_0^1 dX_1\, dY_1 \, dX_5
X_1^{-\alpha' s_{12} - \frac{3}{2}}
(1-X_1)^{-\alpha' s_{23} - \frac{1}{2}}
Y_1^{-\alpha' s_{123} }
(1-Y_1)^{ - \alpha' s_{34} - \frac{1}{2}} 
\nonumber \\
& \times X_5^{ -\alpha' s_{56} - \frac{3}{2}}
(1{-}X_5)^{-\alpha' s_{45} - \frac{1}{2} }
(1{-}X_1 Y_1)^{-\alpha' s_{24}}
(1{-}Y_1 X_5)^{-\alpha' s_{35}} 
(1{-}X_1 Y_1 X_5)^{{-}1{-}\alpha' s_{25} }  \nonumber
\end{align}
The terms $(1-X_1 Y_1)^{-\alpha' s_{24}} $, $(1-X_5 Y_1)^{-\alpha' s_{35}}$ and $(1-X_1 X_5 Y_1)^{-1-\alpha' s_{25}}$ can be expanded using the binomial series. 
The integration over $X_1$, $X_5$ and $Y_1$ is now trivial and we get
\begin{align}
I_{X_1X_5Y_1}= & \sum_{n,\bar{n},m =0}^\infty 
\frac{(-1)^{\bar{n}+n}}{m! n! \bar{n}!}
\frac{\Gamma  (\frac{1}{2}{-}\alpha' s_{23} ) \Gamma  (1{-}\alpha' s_{24} ) \Gamma  (\frac{1}{2}{-}\alpha' s_{34} ) \Gamma  (1{-}\alpha' s_{35} ) \Gamma  (\frac{1}{2}{-}\alpha' s_{45} ) }
{ \Gamma  (1{+}\alpha' s_{25} ) }\times
\nonumber \\
&\times
\frac{ \Gamma  (1{+}m{+}\alpha' s_{25} ) \Gamma  ({-}\frac{1}{2}{+}m{+}n{-}\alpha' s_{12} ) \Gamma  ({-}\frac{1}{2}{+}m{+}\bar{n} {-}\alpha' s_{56}) }
{\Gamma  (\frac{1}{2}{+}m{+}n{+}\bar{n}{-}\alpha' (s_{34}{+} s_{123}))}\times  \\
& \times
\frac{ \Gamma  (m{+}n{+}\bar{n}{-}\alpha' s_{123} )}
{\Gamma  (1{-}n{-}\alpha' s_{24} ) \Gamma  (1{-}\bar{n}{-}\alpha' s_{35} ) \Gamma [m{+}n{-}\alpha' (s_{12}{+}s_{23})] \Gamma [m{+}\bar{n}{-}\alpha'( s_{45}{+} s_{56})] }
\nonumber
\end{align}
In the limit $\a' \to 0$, we can put $\alpha ' =0$ in all the $\Gamma$ functions in the first two lines obtaining constant contributions. 
Those in the  third line instead can yield non-trivial powers of $\alpha'$. The $\Gamma$ function in the  numerator is associated to the massless pole and yields a factor $(\alpha')^{-1}$ when $n{\,=\,}\bar{n}{\,=\,}m{\,=\,}0$.
The four $\Gamma$-functions in the denominator can separately yield a factor of $\alpha '$ in the numerator when the argument of these $\Gamma$ functions vanish. This happens when the following conditions are satisfied:
\begin{itemize}
\item $\Gamma (1{-}n{-}\alpha' s_{2 4})$ when $n=1+N$ (with $N\geq 0$)
\item $\Gamma (1{-}\bar{n}{-}\alpha' s_{3 5})$ when $\bar{n}=1+N$ (with $N\geq 0$)
\item $\Gamma [m{+}n{-}\alpha' (s_{1 2}{+}s_{2 3})]$ when $m{+}n=0$
\item $\Gamma [m{+}\bar{n}{-}\alpha' (s_{4 5}{+}s_{5 6})]$  when $m{+}\bar{n}=0$
\end{itemize}
From the previous considerations, it follows that, in order to obtain the expansion at the order $(\alpha')^1$, we have to focus only on the following regions:
\begin{itemize}
\item $m= n= \bar{n}=0$
\item $m,n>1$ and $\bar{n}=0$
\item $m,\bar{n}>1$ and $n=0$
\item $m>1$ and $n=\bar{n}=0$
\end{itemize}
The contribution of the first region gives immediately 
\begin{equation}
I_{X_1X_5Y_1}^{(I)}=-\frac{4  \alpha ' \pi ^2 \left(s_{12}+s_{23}\right) \left(s_{45}+s_{56}\right)}{s_{123}}+O(\alpha'^2)
\end{equation}
In the second region, using the new variables $N=\bar{n}-1$ and $M=m-1$ and the residue of the $\Gamma$ function at its poles
\begin{equation}
\frac{1}{\Gamma ({-}N{-}\alpha' s_{3 5})}\sim -\alpha' s_{35} ({-}1)^N N!+ O(\alpha'^2) \,\, ,
\end{equation}
we can expand the series in $\alpha'$ obtaining
\begin{align}
I_{X_1X_5Y_1}^{(II)}&= 2 \alpha ' \pi ^{3/2} s_{24} \sum_{N,M=0}^\infty  \frac{  \Gamma \left(M+\frac{1}{2}\right)}{(N+1) (2 N+2 M+3) \Gamma (M+1) }= \nonumber\\
&= 2 \alpha ' \pi ^{3/2} s_{24} \sum_{M=0}^\infty  \frac{\Gamma \left(M+\frac{1}{2}\right) \left[\psi\left(M+\frac{3}{2}\right)+\gamma_E \right]}{(2 M+1) \Gamma (M+1)}= \nonumber \\
&= 2 \pi ^3 \alpha' \, s_{24}\, \log 2 +O(\alpha'^2)
\end{align}
where $\psi(z)=\G'(z)/\G(z)$ is the Digamma function and $\gamma_E=-\G'(1)$ is the Euler-Mascheroni constant. The contribution coming from the third region is similar
\begin{equation}
I_{X_1X_5Y_1}^{(III)}=2 \pi ^3 \alpha' \log 2 \, s_{35}+O(\alpha'^2)
\end{equation}
In the fourth region, as done previously, we introduce the new variable $M=m-1$. The expansion has several terms
\begin{align}
I_{X_1X_5Y_1}^{(IV)}= \sum_{M=0}^\infty &
\Bigg[\frac{ 4 \pi ^{3/2} \alpha '(s_{34}{+}s_{123})\Gamma \left(M{+}\frac{1}{2}\right)}{(2 M{+}1)^2 \Gamma (M{+}1)}+
\frac{2 \pi ^{3/2}  \alpha ' s_{25} \Gamma \left(M{+}\frac{1}{2}\right)}{(2 M{+}1) \Gamma (M{+}2)}+ \\
&+
\frac{2 \pi ^{3/2}  \left[1+\alpha'  s_{25} \gamma_E+ \alpha'(\gamma_E +\log 4) \left(s_{23}{+}s_{34}{+}s_{45}\right) \right] \Gamma \left(M+\frac{1}{2}\right)}{(2 M+1) \Gamma (M+1)}\nonumber \\
&+
\frac{2 \pi ^{3/2}\alpha' \left(s_{25}{+}s_{45}{+}s_{56}{-}s_{13}\right)\Gamma \left(M+\frac{1}{2}\right) \psi(M{+}1)}{(2 M{+}1) \Gamma (M+1)} \nonumber\\
&
+\frac{2 \pi ^{3/2} \alpha' \left(s_{12}{+}s_{34}{+}s_{123}{-}s_{56}\right) \Gamma \left(M+\frac{1}{2}\right) \psi\left(M+\frac{1}{2}\right)}{(2 M+1) \Gamma (M+1)} \Bigg] \nonumber
\end{align}
Summing separately these contributions we obtain
\begin{align}
I_{X_1X_5Y_1}^{(IV)}= & \, \pi ^3{+}
8 \pi ^2  \alpha ' s_{45} G{+}2 \pi ^2 \alpha ' (2s_{13}{-}s_{56})  (\pi  \log 2{-}2 G){+}\\
&{+}2 \pi ^2 \alpha ' s_{25} (4 G{+}\pi {-}2{-}\pi  \log 2) {+}2\pi ^3 \alpha '( s_{12} {+}2s_{23}{+}2s_{34}) \log 2 \nonumber
\end{align}
where $G$ is Catalan constant\footnote{Its explicit expression suggests that $G$ have transcendentality two. We thank Oliver Schlotterer for posing the riddle.}
\begin{equation}
G = \sum_{n=0}^\infty \frac{(-1)^n}{(2n+1)^2}
= \frac{i}{2} \left[
\text{Li}_2\left(\frac{1}{2}-\frac{i}{2}\right)-
\text{Li}_2\left(\frac{1}{2}+\frac{i}{2}\right)\right]
+\frac{\p}{8} \log 2 
\end{equation}
Combining the contributions from all the regions we get
\begin{align}
I_{X_1X_5Y_1}=&\,\pi ^3{-}4 \alpha '\pi ^2 \frac{ \left(s_{12}{+}s_{23}\right) \left(s_{45}{+}s_{56}\right) }{s_{123}}{-}2 \pi ^2\alpha '  \left(s_{16}{+}s_{34}\right) (2{-}4 G{-}\pi {+}\pi  \log 2)\\
&
{-}2 \pi ^2 \alpha '  \left(s_{12}{+}s_{23}{+}s_{45}{+}s_{56}\right) (\pi  \log 2{-}4 G){+}4 \pi ^2  \alpha ' s_{123}  (\pi  \log 2{-}2 G)+\nonumber\\
&{+}2 \pi ^2  \alpha ' \left(s_{234}{+}s_{345}\right)(2{+}2 \pi  \log 2{-}4 G{-}\pi)+O(\ap^2)\nonumber
\end{align}
Notice that the expansion above and the integral in Eq \eqref{example_integral} share the reflection symmetries $1\leftrightarrow 6$, $2\leftrightarrow 5$ and $3\leftrightarrow 4$ of the external legs.

\subsection{Expansions of all the relevant integrals}
\label{expansions_relevant_integrals}

In the previous subsection we have explicitly shown how to proceed in the case of the third term in Eq. (\ref{eq:relevant_integrals}).  One should proceed in a similar way for the other six terms.
Here we just give the result of the $\alpha'$-expansion of the integrals obtained  from  the seven terms in Eq. \eqref{eq:relevant_integrals}:
\begingroup
\allowdisplaybreaks
\begin{align}
\frac{1}{X_1 X_3 X_5}=& {-}\frac{13 \pi ^3}{8}
{+}\ap\frac{\pi ^2}{8} \Big[(s_{16}{+}s_{23}{+}s_{45})({-}104 G{-}23 \pi {+}124{+}26 \pi  \log 2)+\nonumber\\
&{+}(s_{12}{+}s_{34}{+}s_{56}) ({-}104 G{-}33 \pi {+}124{+}26 \pi  \log 2)+\nonumber\\
&
{-}4(s_{123}{+}s_{234}{+}s_{345}) ({-}26 G{-}13 \pi {+}31{+}13 \pi  \log 2)\Big]+O(\alpha'^2) \\
\frac{1}{X_1 X_4 Y_1}=& \,0+O(\alpha'^1)  \label{X1X4Y1_int}\\
%
%
%
%
\frac{1}{X_1 X_5 Y_1}=&\,\pi ^3{-}2 \ap \pi^2 \Big[2\frac{ (s_{12}{+}s_{23}) (s_{45}{{+}}s_{56}) }{s_{123}}{+} (s_{16}{+}s_{34}) (2{-}4 G{-}\pi {+}\pi  \log 2)+\nonumber\\
&
{+} (s_{12}{+}s_{23}{+}s_{45}{+}s_{56}) (\pi  \log 2{-}4 G){-}2 s_{123}  (\pi  \log 2{-}2 G){+}\nonumber\\
&{-}(s_{234}{+}s_{345})(2{+}2 \pi  \log 2{-}4 G{-}\pi)\Big]+O(\alpha'^2)\\
\frac{1}{X_1 Y_1}=&\,0+O(\alpha'^0) \label{X1Y1_int}\\
\frac{1}{X_1}=& \frac{\pi ^3}{2}
{-}\ap \pi ^2 \Big[(s_{16}{+}s_{23}) (\pi \log 2{-}4 G{-}2)
{+}(s_{12}{+}s_{45})( \pi  \log 2{-}4 G{-}\frac{\pi}{2} {-}2){+}~\nonumber\\
&{+}(s_{34}{+}s_{56}) ({-}4 G{+}\pi {-}2{+}\pi  \log 2)
{-}s_{234}  ({-}4 G{+}\pi {-}2{+}2 \pi  \log 2)+\nonumber\\
&
{-}\frac{1}{2}(s_{123}{+}s_{345}) ({-}8 G{+}\pi {-}4{+}4 \pi  \log 2)\Big]+O(\alpha'^2)
\\
1=& \,\pi ^3
{-}2 \ap \pi ^2 \Big[ (s_{12}{+}s_{16}{+}s_{23}{+}s_{34}{+}s_{45}{+}s_{56}) (\pi  \log 2{-}4 G)+\nonumber\\
&{-}2 (s_{123}{+}s_{234}{+}s_{345})  (\pi  \log 2{-}2 G)\Big]+ O(\alpha'^2)\\
X_2 X_4 X_6 Y_1 Y_2 Y_3 =&\frac{\pi ^3}{8} 
{-}\ap \frac{\pi ^2 }{8} \Big[(s_{16}{+}s_{23}{+}s_{45}) ({-}8 G{-}3 \pi {+}12{+}2 \pi  \log 2)+\nonumber\\
&
{+} (s_{12}{+}s_{34}{+}s_{56}) ({-}8 G{-}5 \pi {+}12{+}2 \pi  \log 2)+\nonumber\\
&
{-}4(s_{123}{+}s_{234}{+}s_{345}) ({-}2 G{-}\pi {+}3{+}\pi  \log 2)\Big]+O(\alpha'^2)
\end{align}
\endgroup
Notice that in order to compute the leading term in the $\ap$-expansion of the six-pion amplitude, for the integrals \eqref{X1Y1_int} and \eqref{X1X4Y1_int} it is enough to look at terms with powers of $\ap$ at most $\ap^0$ and $\ap^{-1}$ respectively because the correspondent numerators contain  $\ap^1$ and $\ap^2$ respectively.
We have chosen the (linearly) independent basis $\{s_{i,i+1},s_{i,i+1,i+2}\}$ to show that the integrals and their expansions share the same symmetries. With the expansion of these seven integrals one can easily obtain the expansion of the other integrals shown in Eqs. (\ref{eq:decompositionI1}-\ref{eq:decompositionI3}) using cyclical permutations and reflections. 

Using the expansions of the integrals in Eqs. (\ref{eq:decompositionI1}-\ref{eq:decompositionI3}) in Eq. \eqref{eq:paolo_6pt_XY}, the leading term of the $\ap$-expansion is given by
\begin{align}
A_6&=- \alpha' \pi^2  C_{\textup{6-pt}} \Bigg[\frac{(s_{12}{+}s_{23})(s_{34}{+}s_{45})}{s_{123}}{+}\frac{(s_{23}{+}s_{34})(s_{45}{+}s_{56})}{s_{234}}{+}\frac{(s_{34}{+}s_{45})(s_{56}{+}s_{16})}{s_{345}}{+} \nonumber\\
& {-} (s_{12}{+}s_{23}{+}s_{34}{+}s_{45}{+}s_{56}{+}s_{16})\Bigg]+O(\ap)
\end{align}
that exactly reproduces the analogous amplitude in the non-linear $\s$-model \eqref{eq:6ptNLSM} when $C_{\textup{6-pt}}$ is given by Eq. \eqref{eq:C6asC4}.

\section{Amplitudes extracted from the six-pion amplitude}
\label{ampl_appendix}

In this Appendix we list some of the scattering amplitudes that can be obtained by  factorisation of the full six-pion amplitude. In the following  we will assume that the  flavour group is $U(N_f)$.

Starting from the six-pion amplitude, one can extract the three-point amplitudes involving only $\r$ and/or $\s$ factorizing all the pions in pairs. In this way one obtains:
\begingroup
\allowdisplaybreaks
\begin{align}
A_{\s\s\s}= & -\frac{3\h_\s \sqrt{-C_\textup{4-pt}} }{2  \sqrt{2\ap}} \widehat{d}^{a_{\s_1}a_{\s_2}a_{\s_3}}
\\
A_{\s \s \r}=& -\frac{\h_\r \sqrt{-C_\textup{4-pt}}}{ 2} f^{a_{\s_1}a_{\s_2}a_\r} (p_1-p_2){\cdot}\e_3 \\
A_{\s \r \r}=&\, \h_\s  \frac{\sqrt{-C_\textup{4-pt}}}{\sqrt{2\ap}} \widehat{d}^{a_{\s}a_{\r_1}a_{\r_2}} \e_{2}{\cdot}\e_{3}\\
A_{\s \r \r}=&\, \h_\s  \frac{\sqrt{-C_\textup{4-pt}}}{\sqrt{2\ap}} \widehat{d}^{a_{\s}a_{\r_1}a_{\r_2}} \e_{2}{\cdot}\e_{3}\\
A_{\r\r\r}=& \,2 \h_\r \sqrt{-C_\textup{4-pt}} f^{a_{\r_1}a_{\r_2}a_{\r_3}} 
(\e_{1}{\cdot}\e_{2} \, \e_{3} {\cdot } p_1+\e_2{\cdot}\e_3 \, \e_1 {\cdot } p_2+\e_1{\cdot}\e_3 \, \e_2 {\cdot } p_3)
\end{align}
In the case of the $U(2)$ flavour group, the amplitude $A_{\s \s \r}$ vanishes being $\s$ a singlet.

If we leave one pair of pions factorizing two pairs of pions only, we get the following four-point amplitudes with $\r$ and/or $\s$ and pions:
\begingroup
\allowdisplaybreaks
\begin{align}
A_{2\p 2\s}=
&-\frac{C_\textup{4-pt}}{4} [\Tr(T_1^\p T_2^\p T_3^\s T_4^\s)+\Tr(T_1^\p T_4^\s T_3^\s T_2^\p)] (1{+}2 \ap t) \frac{\Gamma(\frac{1}{2}-\ap s)\Gamma(-\ap t)}{\Gamma(\frac{1}{2}-\ap s- \ap t)}+ \nonumber \\
&-\frac{C_\textup{4-pt}}{4} [\Tr(T_1^\p T_2^\p T_4^\s T_3^\s)+\Tr(T_1^\p T_3^\s T_4^\s T_2^\p)] (1{+}2 \ap u) \frac{\Gamma(\frac{1}{2}-\ap s)\Gamma(-\ap u)}{\Gamma(\frac{1}{2}-\ap s- \ap u)}+ \nonumber\\
&-\frac{C_\textup{4-pt}}{4} [\Tr(T_1^\p  T_3^\s T_2^\p T_4^\s)+\Tr(T_1^\p  T_4^\s T_2^\p T_3^\s)]  \frac{\Gamma(-\ap t)\Gamma(-\ap u)}{\Gamma(-\ap t- \ap u)}
\label{eq:A2pi2sigma}
\end{align}
\begin{align}
A_{2\p 2\r}=
& \ap C_\textup{4-pt}
[\Tr(T_1^\p T_2^\p T_3^\r T_4^\r)+\Tr(T_1^\p T_4^\r T_3^\r T_2^\p)] \times \nonumber\\
&\times \!\Bigg[
(t \e_3{\cdot}\e_4 {-} 2 k_1{\cdot} \e_4 k_2{\cdot} \e_3) \frac{\Gamma(\frac{1}{2}-\ap s)\Gamma(-\ap t)}{\Gamma(\frac{1}{2}-\ap s- \ap t)}+
2 k_2{\cdot} \e_4 k_1{\cdot} \e_3 \frac{\Gamma(\frac{1}{2}-\ap s)\Gamma(1-\ap t)}{\Gamma(\frac{3}{2}-\ap s- \ap t)}
\Bigg]+ \nonumber \\
& +\ap C_\textup{4-pt}
[\Tr(T_1^\p T_2^\p T_4^\r T_3^\r)+\Tr(T_1^\p T_3^\r T_4^\r T_2^\p)] \times \nonumber \\
&\times \!\Bigg[
(u \e_3{\cdot}\e_4 {-}2  k_1{\cdot} \e_3 k_2{\cdot} \e_4) \frac{\Gamma(\frac{1}{2}-\ap s)\Gamma(-\ap u)}{\Gamma(\frac{1}{2}-\ap s- \ap u)}{+}
2 k_2{\cdot} \e_3 k_1{\cdot} \e_4 \frac{\Gamma(\frac{1}{2}-\ap s)\Gamma(1-\ap u)}{\Gamma(\frac{3}{2}-\ap s- \ap u)}
\Bigg]+ \nonumber \\
&+\ap^2 C_\textup{4-pt} [\Tr(T_1^\p  T_3^\r T_2^\p T_4^\r)+\Tr(T_1^\p  T_4^\r T_2^\p T_3^\r)] \times \nonumber \\
&\times (2 u k_1{\cdot} \e_4 k_2{\cdot} \e_3+2 t k_1{\cdot} \e_3 k_2{\cdot} \e_4- tu \e_3 {\cdot}\e_4)\frac{\Gamma(-\ap t)\Gamma(-\ap u)}{\Gamma(-\ap t- \ap u)}
\label{eq:A2pi2rho}
\end{align}
\begin{align}
A_{2\p 1\s 1\r}=
&\frac{i \h_\s \h_\r \sqrt{\ap} C_\textup{4-pt}}{\sqrt{2}}
\Bigg[ [\Tr(T_1^\p T_2^\p T_3^\s T_4^\r){-}\Tr(T_1^\p T_4^\r T_3^\s T_2^\p)]
\e_4 {\cdot}k_1 \frac{\Gamma(\frac{1}{2}-\ap s)\Gamma(-\ap t)}{\Gamma(\frac{1}{2}-\ap s- \ap t)} + \nonumber\\
-& [\Tr(T_1^\p T_2^\p T_4^\r T_3^\s){-}\Tr(T_1^\p T_3^\s T_4^\r T_2^\p)] 
\e_4 {\cdot}k_2 \frac{\Gamma(\frac{1}{2}-\ap s)\Gamma(-\ap u)}{\Gamma(\frac{1}{2}-\ap s- \ap u)}+
\nonumber \\
-&
[\Tr(T_1^\p  T_3^\s T_2^\p T_4^\r){-}\Tr(T_1^\p  T_4^\r T_2^\p T_3^\s)]
(u \e_4{\cdot}k_1{-}t \e_4{\cdot}k_2) \frac{\Gamma(-\ap t)\Gamma(-\ap u)}{\Gamma(1-\ap t- \ap u)} \Bigg]\label{eq:A2pi1sigma1rho}
\end{align}
\endgroup
All these amplitudes inherit the Adler zeroes from the six-pion amplitude. The residues at $\ap s {\,=\,} \frac{1}{2}$ and $\ap t {\,=\,} 0$ are consistent with the three-point amplitudes shown above.

The amplitudes in Eqs. \eqref{eq:A2pi2sigma}, \eqref{eq:A2pi2rho} and \eqref{eq:A2pi1sigma1rho} can be used to find the spectrum at the mass level $m^2=1/\ap$ and the three-point amplitudes involving the lighter particles already found. 
Let us first consider the $\p$-$\s$ scattering amplitude in Eq. \eqref{eq:A2pi2sigma}. The residue at $\ap t =1$ can be reconstructed considering the contribution of one scalar and two vectors whose three-point amplitudes with $\p$ and $\s$ are the followings
\begin{align}
A_{\p \s \tilde{\p}} &= \h_{\tilde{\p}} \frac{3 \sqrt{-C_\textup{4-pt}}}{4 \sqrt{\ap}} \,\widehat{d}_{a_\p a_\s a_{\tilde{\p}}} \\
A_{\p \s a_1} &= \h_{a} \sqrt{-C_\textup{4-pt}}\,  \ve{\cdot}k_\p \,\widehat{d}_{a_\p a_\s a_{a_1}}\\
A_{\p \s h_1} &= \h_{h} \sqrt{-C_\textup{4-pt}}\,  \ve{\cdot}k_\p \,\widehat{f}_{a_\p a_\s a_{h_1}}
\end{align}
where $\h_{\tilde{\p}}$, $\h_{a}$ and $\h_{h}$ are undetermined signs. In the $U(2)$ case the first and second state are isovectors and also their parity is respectively ${-1}$ and ${+1}$. Therefore they might be interpreted as a massive pion $\tilde{\p}$ and a $a_1$ vector meson. The amplitude associated to the third state vanishes in the $U(2)$ case, therefore this state is a singlet. The parity of this state is ${+1}$, thus it can be interpreted as a $h_1$ vector meson.

From the amplitude involving one $\s$, one $\r$ and two pions in Eq. \eqref{eq:A2pi1sigma1rho}, we find the coupling of these three states with $\p$ and $\r$:
\begin{align}
A_{\p \r \tilde{\p}} &= -\h_{\r} \h_{\s}\h_{\tilde{\p}} \sqrt{-\frac{C_\textup{4-pt}}{2}} \,\ve{\cdot} k_\p \widehat{f}_{a_\p a_\r a_{\tilde{\p}}} \\
A_{\p \r a_1} &= -\h_{\r} \h_{\s}\h_{a} \sqrt{-\frac{C_\textup{4-pt}}{2 \ap}}\, \ve_\r{\cdot}\ve_a \,\widehat{f}_{a_\p a_\r a_{a_1}}\\
A_{\p \r h_1} &= -\h_{\r} \h_{\s}\h_{h} \sqrt{-\frac{C_\textup{4-pt}}{4 ap}}\, (\ve_\r{\cdot}\ve_h-4 \ap \ve_\r{\cdot}k_\p \ve_h {\cdot} k_\p )\,\widehat{d}_{a_\p a_\s a_{h_1}}
\end{align}
In the pion-$\r$ scattering amplitude in Eq. \eqref{eq:A2pi2rho}, we find again $\tilde{\p}$ and $a_1$ coupled antisymmetrically to the external states. The $t$-channel contains $h_1$ and also one more vector coupled symmetrically. The residue is reconstructed if this vector couples to the pion and the $\r$ meson as follows
\begin{align}
A_{\p \r \w} & =2  \h_\w \sqrt{-\ap C_\textup{4-pt} } \, \e_{\m_1 \m_2 \m_3 \m_4} \ve_\r^{\m_1} \ve_\w^{\m_2} k_\r^{\m_3} k_\w^{\m_4} \widehat{d}_{a_\p a_\r a_\w}
\end{align}
where $\h_{\w}$ is another undetermined sign. In the $U(2)$ case this state is a singlet and it has parity ${-1}$. Therefore, it might be interpreted as the $\w$ meson. Notice also that the $\w$ meson couples with $\p$ and $\r$ via the $\epsilon^{\mu \nu \rho \sigma}$ symbol as expected.

From the factorization of the six-pion amplitude in the three-pion channel at $\ap s_{123}=1$ we can obtain the scattering amplitude for the processes $X \rightarrow \pi \pi \pi$, where $X$ can be $\tilde{\p}$, $a_1$, $h_1$ and $\w$. The flavour-ordered amplitudes are the following
\begingroup
\allowdisplaybreaks
\begin{align}
A[1_\p,2_\p,3_\p,4_{\tilde{\p}}]&=
-\h_{\tilde{\p}} \h_\s\frac{C_\textup{4-pt}}{2 \sqrt{2}}
\frac{\Gamma (\frac{1}{2}-\ap s) \Gamma(\frac{1}{2}-\ap t)}{\Gamma(-\ap s -\ap t)} \\
A[1_\p,2_\p,3_\p,4_{a}]&=
-\h_{a} \h_\s\frac{C_\textup{4-pt} \sqrt{\ap}}{\sqrt{2}} \ve_a{\cdot}k_2
\frac{\Gamma (\frac{1}{2}-\ap s) \Gamma(\frac{1}{2}-\ap t)}{\Gamma(1-\ap s -\ap t)}\\
A[1_\p,2_\p,3_\p,4_{h}]&=
-i \h_{h} \h_\s C_\textup{4-pt} \ap^{3/2}(s \ve_h{\cdot}k_1-t \ve_h{\cdot}k_3)
\frac{\Gamma (\frac{1}{2}-\ap s) \Gamma(\frac{1}{2}-\ap t)}{\Gamma(1-\ap s -\ap t)}\\
A[1_\p,2_\p,3_\p,4_\w]&=
2 i \h_\w \h_\r C_\textup{4-pt} \ap^{3/2}
\epsilon^{\mu \nu \rho \sigma} \ve_\mu k_{1 \nu }k_{2 \rho} k_{3 \sigma}
\frac{\Gamma (\frac{1}{2}-\ap s) \Gamma(\frac{1}{2}-\ap t)}{\Gamma(1-\ap s -\ap t)}
\end{align}
\endgroup
As for the amplitudes in Eqs. \eqref{eq:A2pi2sigma}, \eqref{eq:A2pi2rho} and \eqref{eq:A2pi1sigma1rho}, these four amplitudes inherit the Adler zeroes from the six-pion amplitude. The residues at $\ap s {\,=\,} {1}/{2}$ are consistent with the three-point amplitudes shown above.



\begin{thebibliography}{10}

\bibitem{Veneziano:1968yb}
G.~Veneziano,
``Construction of a crossing - symmetric, Regge behaved amplitude for 
linearly rising trajectories,''
\href{https://doi.org/10.1007/BF02824451}{Nuovo Cim.\ A {\bf 57} (1968) 190}.

\bibitem{LOVELACE}
C.~Lovelace,
``A novel application of Regge trajectories,''
\href{https://doi.org/10.1016/0370-2693(68)90255-4}
{Phys.\ Lett.\  {\bf 28B}, 264 (1968)}.

\bibitem{SHAPIRO}
J.~A.~Shapiro,
``Narrow-resonance model with Regge behavior for pi pi scattering,''
\href{https://doi.org/10.1103/PhysRev.179.1345}
{Phys.\ Rev.\  {\bf 179}, 1345 (1969)}.

\bibitem{FRAMPTON}
P.~H.~Frampton,
``O(n) relations for coupling constants and space-time dimensions in dual models,''
\href{https://doi.org/10.1016/0370-2693(72)90597-7}
{Phys.\ Lett.\  {\bf 41B}, 364 (1972)}.

\bibitem{PDV}
P. Di Vecchia,
``The birth of string theory'',
Lect.\ Notes Phys.\  {\bf 737}, 59 (2008)
 \hri{0704.0101}{[hep-th]}.

\bibitem{PR}
P.~Ramond,
``Dual Theory for Free Fermions,''
\href{https://doi.org/10.1103/PhysRevD.3.2415}
{Phys.\ Rev.\ D {\bf 3}, 2415 (1971)}.

\bibitem{NS}
A.~Neveu and J.~H.~Schwarz,
``Factorizable dual model of pions'',
\href{https://doi.org/10.1016/0550-3213(71)90448-2}
{Nucl.\ Phys.\ B {\bf 31}, 86 (1971)}.

\bibitem{NST}
A.~Neveu, J.~H.~Schwarz and C.~B.~Thorn,
``Reformulation of the Dual Pion Model,''
\href{https://doi.org/10.1016/0370-2693(71)90391-1}
{Phys.\ Lett.\  {\bf 35B}, 529 (1971)}.

\bibitem{FM}
D.~B.~Fairlie and D.~Martin,
``New light on the Neveu-Schwarz model,''
\href{https://doi.org/10.1007/BF02722834}
{Nuovo Cim.\ A {\bf 18}, 373 (1973)}.

\bibitem{BW}
L.~Brink and J.-O.~Winnberg,
``The Superoperator Formalism of the Neveu-Schwarz-Ramond Model,''
\href{https://doi.org/10.1016/0550-3213(76)90509-5}
{Nucl.\ Phys.\ B {\bf 103}, 445 (1976)}.

\bibitem{KH}
K.~Hornfeck,
``Three Reggeon Light Cone Vertex of the Neveu-schwarz String,''
\href{https://doi.org/10.1016/0550-3213(87)90068-X}
{Nucl.\ Phys.\ B {\bf 293}, 189 (1987)}.

\bibitem{CP}
Jack E. Paton and Hong-Mo Chan,
``Generalized veneziano model with isospin,''
\href{https://doi.org/10.1016/0550-3213(69)90038-8}
{Nucl. Phys. {\bf B10} (1969) 516}.

\bibitem{BrowerPLB71} 
R.~C.~Brower,
``A chiral invariant dual model,''
\href{https://doi.org/10.1016/0370-2693(71)90691-5}
{Phys.\ Lett.\  {\bf 34B}, 143 (1971)}.

\bibitem{NeveuThornPRL71} 
A.~Neveu and C.~B.~Thorn,
``Chirality in dual resonance models,''
\href{https://doi.org/10.1103/PhysRevLett.27.1758}
{Phys.\ Rev.\ Lett.\  {\bf 27}, 1758 (1971)}.

\bibitem{SchwarzPRD72}
J.~H.~Schwarz,
``Dual-pion model satisfying current-algebra constraints,''
\href{https://doi.org/10.1103/PhysRevD.5.886}
{Phys.\ Rev.\ D {\bf 5}, 886 (1972)}

\bibitem{FairlieNPB72} 
D.~B.~Fairlie,
``An empirical extension of the Neveu-Schwarz model,''
\href{https://doi.org/10.1016/0550-3213(72)90478-6}
{Nucl.\ Phys.\ B {\bf 42}, 253 (1972)}.

\bibitem{Carrasco:2016ldy}
J.J. Carrasco, C.R. Mafra and O. Schlotterer,
``Abelian Z-theory: NLSM amplitudes and $\alpha$'-corrections from the open string,''
JHEP {\bf 06} (2017) 093, 
\hri{1608.02569}{[hep-th]}.

\bibitem{BrowerChuPRD73} 
R.~Brower and G.~Chu,
``Phenomenological six-pion amplitude,''
\href{https://doi.org/10.1103/PhysRevD.7.56}
{Phys.\ Rev.\ D {\bf 7}, 56 (1973)}.
  
\bibitem{SinghPasupathyPRD74} 
C.~A.~Singh and J.~Pasupathy,
``Phenomenological, dual, multipion amplitudes,''
\href{https://doi.org/10.1103/PhysRevD.10.1655}
{Phys.\ Rev.\ D {\bf 10}, 1655 (1974)}.

\bibitem{CKSZ}
S. Caron-Huot,  Z. Komagodski, A. Sever and A. Zhibodeov,
``Strings from massive higher spins: the asymptotic uniqueness of the Veneziano amplitude'',
JHEP {\bf 1710} (2017) 026,
 \hri{1607.04253}{[hep-th]}.

\bibitem{AS}
O. Andreev and W. Siegel,
``Quantized tension: Stringy amplitudes with Regge poles and parton behavior,''
Phys. Rev D {\bf 71} (2005) 086001,
 \hre{hep-th}{0410131}.

\bibitem{VYO}
G. Veneziano, S. Yankielowicz and E. Onofri,
``A model for pion-pion scattering in large-N QCD``,
JHEP {\bf 1704} (2017) 151,
 \hri{1701.06315}{[hep-th]}.

\bibitem{Witten:1998zw} 
E.~Witten,
``Anti-de Sitter space, thermal phase transition, and confinement in gauge theories``,
Adv.\ Theor.\ Math.\ Phys.\  {\bf 2}, 505 (1998),
 \hre{hep-th}{9803131}.
  
\bibitem{Sakai:2004cn} 
T.~Sakai and S.~Sugimoto,
``Low energy hadron physics in holographic QCD,''
Prog.\ Theor.\ Phys.\  {\bf 113}, 843 (2005),
 \hre{hep-th}{0412141}.

\bibitem{Bartolini:2016dbk} 
L.~Bartolini, F.~Bigazzi, S.~Bolognesi, A.~L.~Cotrone and A.~Manenti,
``Theta dependence in Holographic QCD,''
JHEP {\bf 1702}, 029 (2017),
 \hri{1611.00048}{[hep-th]}.

\bibitem{Aharony:2006da} 
O.~Aharony, J.~Sonnenschein and S.~Yankielowicz,
``A Holographic model of deconfinement and chiral symmetry restoration,''
Annals Phys.\  {\bf 322}, 1420 (2007),
 \hre{hep-th}{0604161}.

\bibitem{Sonnenschein:2016pim} 
J.~Sonnenschein,
``Holography Inspired Stringy Hadrons,''
Prog.\ Part.\ Nucl.\ Phys.\  {\bf 92}, 1 (2017),
 \hri{1602.00704}{[hep-th]}.
  
\bibitem{Sonnenschein:2017ylo} 
J.~Sonnenschein and D.~Weissman,
``The decay width of stringy hadrons,''
Nucl.\ Phys.\ B {\bf 927}, 368 (2018),
 \hri{1705.10329}{[hep-th]}.

\bibitem{Sonnenschein:2018fph} 
J.~Sonnenschein and D.~Weissman,
``Excited mesons, baryons, glueballs and tetraquarks: Predictions of the Holography Inspired Stringy Hadron model,''
Eur.\ Phys.\ J.\ C {\bf 79}, no. 4, 326 (2019),
 \hri{1812.01619}{[hep-th]}.
 
\bibitem{Sonnenschein:2019bca} 
J.~Sonnenschein, D.~Weissman and S.~Yankielowicz,
``The scattering amplitude of stringy hadrons I: Strings with opposite charges on their endpoints,''
 \hri{1906.00976}{[hep-th]}.
    
\bibitem{Guerrieri:2019rwp} 
A.~L.~Guerrieri, J.~Penedones and P.~Vieira,
``Bootstrapping QCD Using Pion Scattering Amplitudes,''
Phys.\ Rev.\ Lett.\  {\bf 122}, no. 24, 241604 (2019),
 \hri{1810.12849}{[hep-th]}.

\bibitem{probing}
M.~Bianchi, D.~Consoli and J.~F.~Morales,
``Probing Fuzzballs with Particles, Waves and Strings,''
JHEP {\bf 1806} (2018) 157,
 \hri{1711.10287}{[hep-th]}.

\bibitem{HT}
M.~B.~Halpern and C.~B.~Thorn,
``Dual model of pions with no tachyon,''
\href{https://doi.org/10.1016/0370-2693(71)90416-3}
{Phys.\ Lett.\  {\bf 35B}, 441 (1971)}.

\bibitem{BPST}
R.C. Brower, J. Polchinski, M.J. Strassler and C. I. Tan,
``The Pomeron and gauge/string duality,''
JHEP {\bf  0712}  (2007) 005,
 \hre{hep-th}{0603115}.

\bibitem{AI}
A. Armoni and E. Ireson,
``Holographic Corrections to Meson Scattering Amplitudes,''
Nucl.\ Phys.\ B {\bf 919}, 238 (2017),
 \hri{1611.00342}{[hep-th]}.
 
\bibitem{Mafra:2011nw} 
C.~R.~Mafra, O.~Schlotterer and S.~Stieberger,
``Complete N-Point Superstring Disk Amplitude II. Amplitude and Hypergeometric Function Structure,''
Nucl.\ Phys.\ B {\bf 873}, 461 (2013),
 \hri{1106.2646}{[hep-th]}.

\end{thebibliography}
\end{document}